\documentclass[12pt]{article}%

\usepackage[
  a4paper, mag=1000,
  left=2.25cm, right=2.25cm, top=2cm, bottom=2cm
]{geometry}
\newcommand\myspacing{\renewcommand\baselinestretch{1.0}\normalsize}
\myspacing

\usepackage{amssymb}
\usepackage{amsfonts}
\usepackage{amsmath}%
\usepackage{graphicx}

\begin{document}

\title{Transformational process zone emerging at the tip of a propagating crack}
\author{A. Boulbitch\\IEE S.A. ZAE Weiergewan, 11, rue Edmond Reuter, \\L-5326 Contern, Luxembourg;
\and A. L. Korzhenevskii\\Institute for Problems of Mechanical Engineering, \\RAS, Bol'shoi prosp. V. O. 61, \\199178 St. Petersburg, Russia}
\maketitle

\begin{abstract}
Process zone at the tip of a propagating crack engendered by the
stress-induced local phase transition of the second order is studied
theoretically. We show that the zone can only exist within a certain domain of
the phase diagram at one side of the phase transition line depending upon the
sign of the striction constant. We obtain the boundary of this domain and
establish its dependence upon the crack velocity. We show the existence of \ a
critical crack velocity above which the zone cannot exist. We report the
overcritical solution for the order parameter describing the incipient process
zone, while far from the bifurcation point we solve the problem numerically.

\end{abstract}

\section{Introduction}

It is important to study the brittle fracture both from the point of view of
numerous applications and in order to further propel our knowledge of
intrinsic properties of solids. A set of experimental facts presently
collected on fracture is very impressive. However, the understanding of
physical aspects of the its mechanisms is still far from completeness.

Since decades a common opinion worked out that the crack behavior is
determined by the process zone, a nano- to mesosized domain in the vicinity of
its tip. For the long time, however, the process zone was beyond the reach of
experiments. Recently new experimental techniques (described in more details
below) emerged that enable one to study structure and properties of the solid
in the close vicinity of the crack tip. These as well as some more traditional
experimental techniques have shown that fracture is often followed by a
rearrangement of the solid structure in the close vicinity of the crack tip.
This challenges one to find a theoretical approach adequately describing such
a phenomenon.

Stress-induced local phase transitions (LPTs) at crack tips have been reported
in literature since long time. They have been observed in different classes of materials.

Stress induced austenite-martensite LPTs in metals have been studied since
60th \cite{Antolovich Hornbogen} and are under a keen attention up to now.
Martensite LPTs have been observed in iron \cite{iron} and steels
\cite{steels}, \cite{Roth}. Active studies of the martensite LPT in Ni-Ti
(Nitinol) alloys widely-used in application are being actively carried out
\cite{Nitinol}, \cite{Robertson}, \cite{Daly}. LPTs are also exhibited by
other shape-memory alloys, such as Cu-Al-Ni \cite{CuAlNi}, \cite{Lu}, Cr-Ni
\cite{CrNi}, Ni-Al \cite{NiAl} and Ti-Al-Nb \cite{TiAlNb}.

Crack tip LPT from the bulk bcc phase into the nanoscale fcc phase zone has
been very recently observed in molibdenum \cite{molibdenum}, the latter phase
being nonexistent under pressure, $p\geq0$.

ZrO$_{2}$ based ceramics have received a large attention, since it has been
observed that the tetragonal-monoclinic phase transition at the crack tip
strongly improves their fracture toughness \cite{Birkby and Stevens},
\cite{Kelly and Rose} \cite{Todd and Saran}. Ferro- and antiferroelectric
ceramics have also been reported to exhibit LPT under fracture
\cite{ferroelectrics}, \cite{Raman mapping}.

The fracture toughness improvement due to the superconducting LPT in YBCO and
BSCCO at the crack tip has been reported in the paper \cite{Supercond}.

A structural rearrangement within the crack tip zone in sapphire manifested in
formation of metastable Al-O-Al clusters at its fracture surface has been
recently reported \cite{sapphire}.

The crack tip stress-induced structural LPTs have been also observed in
polymers and epoxies \ \cite{polymers}. Resins are known to exhibit
crystallization at the crack tip strongly affecting the fracture process
\cite{resins}.

As for general trends in the evolution of the methods of the LPT observation
one can note a transition from indirect methods of analysis to those making it
possible to obtain a direct structure of the transformed zone with a high
spatial resolution, combining few techniques in one study often appropriate
for propagating cracks. One should first of all mention the method of
high-angle annular dark-field scanning TEM approach allowing for the direct
imaging of atomic locations. It is this method that has been recently used to
detect the crack tip LPT in molibdenum. To exclude any interpretation
ambiguity the results have been further combined with the electron
nanodiffraction patterns \cite{molibdenum}. Further, the combination of
micromechanical loading with \textit{in-situ} high-resoluion X-ray
microdiffraction \cite{Robertson} enabled the authors to image a complex LPT
zone in the polycrystalline Nitinol. The combination of the \textit{in-situ}
SEM with the electron backscatter diffraction \cite{Roth} made it possible to
study the evolution of the emerging phase in the tip vicinity during the
fatigue experiments. The \textit{in-situ} optical digital image correlation
technique made it possible to obtain strain fields and the phase boundaries at
the tips of propagating cracks \cite{Daly}. Raman mapping revealed the local
distribution of phases in the vicinity of the crack tip \cite{Raman mapping}.
AFM maps the lateral distribution of the surface height. The latter is
directly related to spontaneous phase transition strain, thus, enabling one to
distinguish phases and determine the phase boundary \cite{Meschke}, \cite{Lu}.

On the theoretical side three approaches can be pointed out. First, atomistic
mechanisms of LPTs have been revealed by computer simulations and density
functional theory-like calculations for several solids, such as iron
\cite{Nishimura}, \cite{Guo}, \cite{iron simulations}, silicon \cite{Buehler},
\cite{Sherman1}, \cite{Sherman2}, tantalum \cite{Ta simulations}, zirconium
\cite{Zr simulation}, UO$_{2}$ \cite{UO2 simulation}, molibdenum
\cite{molibdenum}, Nitinol \cite{NiTi simulation}. This became possible as the
result of the computation power development, and advancement of the molecular
dynamics approach, as well as in implementation of hybrid approaches combining
the molecular dynamics with quantum mechanics \cite{Sherman2}. The simulations
revealed a strong dependence of the LPT formation upon (i) loading mode, (ii)
crack plane and direction and (iii) sample geometry \cite{Nishimura},
\cite{Guo}. They further elucidated atomistic mechanisms leading to the LPT
development \cite{Guo}, \cite{Sherman2}.

Second, a number of researches exploited a mechanical approach treating the
LPT zone as the one only differing from the rest of the solid by its (i)
elastic properties and (ii) spontaneous strain. Antolovich \cite{Antolovich}
was first to propose a mechanism of the transformation toughness of a
quasi-static crack. References to further papers of this kind one finds in the
reviews \cite{Birkby and Stevens}, \cite{Kelly and Rose} \cite{Todd and Saran}
as well as in the book \cite{Karihaloo}.

The above approach ignores the fact that the local phase transition is related
to one or several internal degrees of freedom of the solid obeying their
intrinsic constitutive laws. A third stream of works just focused on the role
of the internal degrees of freedom in the formation of the LPT at extended
defects. It has been pioneered by the paper of Nabutovsky and Shapiro
describing the dislocation-induced LPT \cite{Nabutovskii and Shapiro}. In the
papers of Korzhenevskii the effect of the LPT on the behavior of dislocations
and, thus, on plastic properties of the solid has been established
\cite{Korzhenevskii},\label{A Now} formation of the LPT at a moving
dislocation has been described in the paper \cite{Boulbitch and Pumpjan}. LPT
at wide domain walls has been described in \cite{Bulbich and Gufan}, while
those at narrow twin boundaries has been addressed in the papers \cite{Narrow
DW}. This approach has also been developed in application to brittle fracture.
Formation of the LPT at the crack tip has been analytically described in the
papers \cite{Bulbich} and \cite{Boulbitch and Toledano}, and numerically in
\cite{Levitas} and \cite{Bjerken}.

It is generally accepted that mechanisms of the brittle solid resistance to
the crack propagation are attributed to its process zone (PZ). The PZ notion
only has sense, if such a zone can be clearly distinguished from the bulk of
the solid. The latter can only be done,\ if at least one
its\ physical\ property\ exhibits a perceptible variation
across\ the\ PZ\ boundary.\ It may either be an abrupt quantitative variation
such as the elastic nonlinearity \cite{Bouchbinder} or hyperelasticity
\cite{Gao}, or any qualitative variation. In the latter case the PZ differs
from the bulk by e.g. its chemical composition or crystal structure. Such
properties determining qualitative differences are always controlled\ by
internal solid\ degrees\ of\ freedom, $\eta$, as e.g., concentration of
reaction species in the former and phonons, magnons, electronic degrees of
freedom, etc. in the latter case.

We focus on the case of the PZ qualitatively different form the solid bulk by
its structure described. Condensation of any such structural degree of freedom
in the bulk of the solid ($\eta=const\neq0$) corresponds to a bulk phase
transition (PT): structural, magnetic, electronic respectively. The degree of
freedom related to the condensate, $\eta$, is referred to as the "order parameter".\

High values of stress at the crack tip: $\mathbf{\sigma}\sim r^{-1/2}$ may
trigger formation of a transformational PZ, where $\eta=\eta(\mathbf{r})\neq0
$ within the zone, while vanishing \ outside. This situation
may,\ thus,\ be\ regarded\ as\ a LPT. Here $\mathbf{\sigma}$ is the stress
tensor and $\mathbf{r}$ is the radius-vector counted off from the crack tip
\cite{Cherepanov}.

The present paper reports the case of the second order LPT admitting a fully
analytical treatment. The latter is of a paramount importance, since it gives
hints of what can be expected in more complex cases that cannot be treated analytically.

\begin{figure}
[ptb]
\begin{center}
\includegraphics[
height=1.0274in,
width=3.333in
]%
{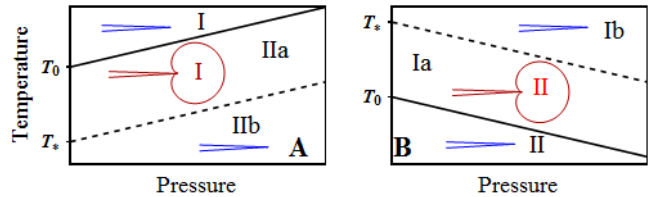}%
\caption{Schematic view of the location of the process zone on the phase
diagram depending upon the slopes of the transition lines. (A) The bulk phases
I and II are separated by the phase transition line, $T_{0}(p)$ (the solid
line), with $k>0$. The region of the process zone existence, IIa, is below the
transition line, where the phase I process zone is embedded into the matrix of
the bulk phase II. In the region IIb, below the line $T_{\ast}(p)$ (the dashed
line) the process zone vanishes. (B) The case of the negative slope, $k<0$, of
the bulk phase transition line, $T_{0}(p)$ (solid). The process zone existence
region, Ia, is situated above the phase transition line, where the phase II is
embedded into the matrix of the phase I. In the region Ib above the $T_{\ast
}(p)$ line (dashed) the zone vanishes.}%
\label{fig1}%
\end{center}
\end{figure}

We show here that in a solid undergoing a second order phase transition a LPT
zone with the size
\begin{equation}
L_{f}\sim10\left(  \frac{g}{a|k|K_{I}}\right)  ^{2/3}\label{Sz}%
\end{equation}
forms at the tip of the motionless as well as propagating crack either above,
or below the line of the phase transition on the phase diagram. We show that
in the both cases the difference, $T_{\ast}-T_{\text{c}}$, between the
transformational process zone (TPZ) emerging temperature, $T_{\ast}$, and the
bulk transition temperature (the Curie point), $T_{\text{c}}$, is
\begin{equation}
T_{\ast}-T_{c}\sim\pm\left(  \frac{a}{g}\right)  ^{1/3}\left(  |k|K_{I}%
\right)  ^{4/3}\mp\frac{\kappa^{2}V^{2}}{4ga}\text{\ \ }\label{DeltaTMain}%
\end{equation}
The difference, $\Delta T_{\ast}=T_{\ast}-T_{c}$, we refer to as the
"temperature shift". Here $K_{I}$ is the stress intensity factor applied to
the crack, $E$ is the Young's modulus, $k=dT_{c}/dp$ is the slope of the phase
transition line in the $(p,T)$ phase diagram, $\kappa$ is the order parameter
kinetic constant and $g$ is the one defining the energy of the order parameter
inhomogeneity, the parameter $a$ is related to the Curie constant, $C$ as
$a=2\pi/C$. The upper sign corresponds to $k<0$, while the lower one - to
$k>0$. Therefore, if the phase diagram slope is positive ($k>0 $), the LPT
only takes place below the phase transition line in the phase diagram $\Delta
T_{\ast}<0$, while the negative slope ($k<0$) results in $\Delta T>0$, that
is, the zone containing a low-temperature phase is embedded into the matrix of
a high-temperature phase (\ref{fig1}).%

We, further, demonstrate the existence of the critical velocity:
\begin{equation}
V_{c}\sim\frac{g^{1/3}}{\kappa}\left(  a\left\vert k\right\vert K_{I}\right)
^{2/3}\label{VcMain}%
\end{equation}
such as soon as the crack tip velocity exceeds the critical one, the TPZ vanishes.

In the discussion we first illustrate the results using the examples of few
ferroelectrics: BaTiO$_{3}$, PbTiO$_{3}$ and LiNbO$_{3}$. We, further,
estimate typical values of $\Delta T_{\ast}$, valid for most inorganic solids.
This represents our most striking result. Indeed, the typically $\Delta
T_{\ast}$ values lie between $\sim10^{2}$ and $\sim10^{3}%
\operatorname{K}%
$. This implies that the TPZ existence region covers a considerable part, if
not the whole phase diagram, though only at one side of the phase transition
line. We then discuss experimental difficulties of the TPZ detecting, as well
as possible ways of its observation. We, finally, put forward a general
approach for the concept of the process zone.

The paper is organized as follows. In Section II we formulate an equation
describing the order parameter dynamics. Lengthily details of its derivation
we give in Appendix A. In Section III we find the analytical solution of the
PZ equations. In Section VI we report our numerical results. In Section V we
make numerical estimates and give comments generalizing our findings.

\section{Process zone dynamics \label{Section II}}

Dynamic behavior of the order parameter splits into several universality
classes as it is suggested by the Hohenberg-Halperin-Ma scheme
\cite{HohenbergHalperinMa}. Within this scheme the most simple and
simultaneously most often met case is referred to as the model A describing
the dynamics of a nonconserved order parameter, $\eta$, corresponding to a
phase transition.

We focus on the case related to crystal structure variation within the process
zone, that is, the case of a structural LPT. For PTs of such a type the order
parameter generally represents a multicomponet object, $\mathbf{\eta}%
=(\eta_{1},\eta_{2},...,\eta_{n})$, each component being composed of
combinations of displacements of the unit crystal cell atoms. Its properties
are completely determined by the corresponding irreducible representation of
the crystal symmetry group in the high-symmetry phase of the solid
\cite{LandauStat}, \cite{Gufan}, \cite{Toledano}. In the high-symmetry phase
$\mathbf{\eta=}0$, while in the bulk low-symmetry phases the components
$\eta_{i}$ ($i=1,2,..$) of the order parameter, are either equal to constants,
or to zero. They are independent of spatial coordinates, at least within a
single domain of a given phase. Such an approach introduced by Landau in 1937
\cite{LandauStat} enables one to classify all phases and to predict possible
bulk phase diagrams \cite{Gufan}, \cite{Toledano}. Here we adapt this approach
for the description of the local transformational PZ.

In this paper we only consider the case of a single-component order parameter
($n=1$). Being simple it already catches the most important properties of the
dynamic LPTs. It is also important for another reason. One observes that in
solids with PTs described by multicomponent order parameters, most of their
bulk low-symmetry phases are described effectively by one independent
parameter. For example, the low-symmetry phases in BaTiO$_{3}$ are generally
described by the three-component polarization vector $\mathbf{\eta=}(\eta
_{1},\eta_{2},\eta_{3})$. However, in the tetragonal phase one finds
$\mathbf{\eta=}(0,0,\eta)$, in the rhombohedric phase - $\mathbf{\eta=}%
(\eta,\eta,\eta)$ and in the orthorhombic phase - $\mathbf{\eta=}(\eta
,\eta,0)$: all these phases are, thus, effectively single-component ones. This
situation is characteristic also for structural PTs in many other materials.
This makes the case of the single-component order parameter distinguished.

\subsection{Equation of motion for the order parameter}

Equation describing the order parameter dynamics can be obtained with the help
of the dissipation function:%
\begin{equation}
D=\frac{\kappa}{2}\int\left(  \frac{\partial\eta}{\partial t}\right)
^{2}d\Omega\label{DissipationFunction}%
\end{equation}
and the free energy:%

\begin{equation}
F=%
{\displaystyle\int}
\Phi(\eta,\varepsilon_{ik})d\Omega\label{FreeEnergy}%
\end{equation}
where $\kappa$ is the kinetic constant, $t$ is the time, $\Phi=\Phi
(\eta,\varepsilon_{ik})$ is the free energy density, $\varepsilon_{ik}$ is the
strain tensor and $\Omega$ is the domain. Since we consider a thin plate case
here, $\Omega$ represents a plane and $d\Omega\equiv dxdy$. That is, we assign
both $F$ and $D$ to the unit solid thickness in the $z$ direction.

Here we address the simplest case of a PT describing by a one-component order
parameter. In the one-component case the only possible transformations under
the action of the crystal symmetry group are either $\eta\rightarrow\eta$ or
$\eta\rightarrow-\eta$. Being invariant with respect to the crystal symmetry,
the free energy should only contain even functions with respect to $\eta$
\cite{LandauStat}. It takes the following form:
\begin{equation}
\Phi(\eta,\varepsilon_{ik})=\Phi_{\text{pt}}(\eta)+\Phi_{\text{el}%
}(\varepsilon_{ik})+A\eta^{2}\varepsilon_{ii}\label{FreeEnergyDensity}%
\end{equation}
Here $\Phi(\eta,\varepsilon_{ik})$ is the free energy density, the function
$\Phi_{\text{pt}}(\eta)$ denoting its part responsible for the PT itself:%

\begin{equation}
\Phi_{\text{pt}}=\frac{g}{2}\left(  \nabla\eta\right)  ^{2}+\frac{\alpha}%
{2}\eta^{2}+\frac{\beta_{0}}{4}\eta^{4}\label{fch}%
\end{equation}
where $g>0$, and $\beta_{0}>0$ are the constant parameters of the Landau
potential (\ref{FreeEnergy}), while $\alpha=a(T-T_{\text{c}})$, where $a>0$ is
a constant, $T$is the temperature and $T_{c}$ is the Curie temperature and
$\nabla\eta$ is the order parameter gradient. The case of the first order
transition $\beta_{0}<0$ will be analyzed elsewhere. The strain tensor,
$\varepsilon_{ik}$, is defined in a usual way:
\[
\varepsilon_{ik}=\frac{1}{2}(\frac{\partial u_{i}}{\partial x_{k}}%
+\frac{\partial u_{k}}{\partial x_{i}})
\]
Here $u_{i}$ is the displacement vector.

$\Phi_{\text{el}}(\varepsilon_{ik})$ is the elastic part of the free energy
density. For simplicity we consider here the elastically-isotropic case with%

\begin{equation}
\Phi_{\text{el}}=\frac{\lambda}{2}\varepsilon_{ii}^{2}+\mu\varepsilon_{ik}%
^{2}\label{fel}%
\end{equation}
where $\lambda$ and $\mu$ are Lame constants \cite{LandauMech}, where%

\begin{equation}
\lambda=\frac{E\nu}{(1-2\sigma)(1+\sigma)};\ \ \mu=\frac{E}{2(1+\sigma
)}\label{LameConst}%
\end{equation}
yields their relations to the Young's modulus and Poisson's ratio, $\sigma$.
The latter should not be confused with the stress tensor $\mathbf{\sigma
}\equiv\sigma_{ik}$. It should be mentioned that the constants $g$, $a$,
$\beta$, $T_{c}$ and $A$ represent the material constants of the solid in
question together with $E$ and $\sigma$.

Finally, the, so-called, striction constant, $A$, already introduced above is
responsible for the interaction between the strain and order parameter fields
and can be either positive or negative. It should be noted that the form of
the interaction term $A\eta^{2}\varepsilon_{ii}$ in (\ref{FreeEnergyDensity})
implies that the phase transition only gives rise to the spontaneous
dilatation. It is only this case that is considered in the present paper. Our
final results are more convenient to express in terms of the slope of the
phase transition line, $k=dT_{c}/dp$, on the $(p,T)$ phase diagram directly
related to the striction constant $A$. Indeed, one can represent the term
$\sim\eta^{2}$ in (\ref{FreeEnergyDensity}, \ref{fch}) as $a\left(
T-T_{c}+Aa^{-1}\varepsilon_{ii}\right)  \eta^{2}$, yielding
\begin{equation}
k=\frac{A(1-2\sigma)}{aE}\label{dT/dp}%
\end{equation}

Equation of motion can be built on the basis of (\ref{FreeEnergy},
\ref{DissipationFunction}) as follows \cite{LandauStat}:
\begin{equation}
\frac{\delta D}{\delta\eta}=-\frac{\delta F}{\delta\eta}\text{; \ \ \ }%
\frac{\delta F}{\delta\varepsilon_{ik}}=0\label{Variation}%
\end{equation}
where $\delta$ is the variation sign. One obtains the following system of
equations:
\begin{equation}%
\genfrac{\{}{.}{0pt}{}{\kappa\frac{\partial\eta}{\partial t}=g\Delta
\eta-[\alpha-2A\varepsilon_{ii}\left(  \mathbf{r}\right)  ]\eta-\beta_{0}%
\eta^{3}}{\partial\sigma_{ik}/\partial x_{k}=0}%
\label{GLK1}%
\end{equation}
where $\Delta$ is the Laplace operator and $\sigma_{ik}=\partial\Phi
/\partial\varepsilon_{ik}$ is the stress tensor:
\begin{equation}
\sigma_{ik}=\lambda\varepsilon_{jj}\delta_{ik}+2\mu\varepsilon_{ik}-A\eta
^{2}\delta_{ik}\label{stress2}%
\end{equation}
Here $\delta_{ik}$ is the Kronecker symbol. The last term, $A\eta^{2}%
\delta_{ik}$, in the expression (\ref{stress2}) describes the spontaneous
stress generated by the phase transition. For simplicity we omitted the
inertial term in the second Eq. (\ref{GLK1}), which is valid, if $V<<c$, where
$c$ is the sound speed.

Equations (\ref{GLK1}, \ref{stress2}) represent the complete system describing
the dynamics of the transformational PZ.

One can eliminate the elastic degrees of freedom, $\varepsilon_{ij}$, form the
equations of motion (\ref{GLK1}) as it is described in details in Appendix A.
After their elimination one comes to the single equation of motion or the
order parameter:
\begin{equation}
\kappa\frac{\partial\eta}{\partial t}=g\Delta\eta-[\alpha+2A\varepsilon
_{ii}^{(0)}\left(  \mathbf{r}\right)  ]\eta-\beta\eta^{3}\label{EffectiveGLK}%
\end{equation}
In contrast to the strain tensor $\varepsilon_{ii}\left(  \mathbf{r}\right)  $
met in Eq. (\ref{GLK1}) describing both the field of the tip and that
generated \ the PZ, the tensor $\varepsilon_{ik}^{(0)}(\mathbf{r})$ only
describes the strain field of the "undressed" tip, i.e., the one without the
LPT ($\eta\equiv0$). It is given by the the well-known fracture theory
expression \cite{Cherepanov}:
\begin{equation}
\varepsilon_{ii}^{(0)}(\mathbf{r})=\frac{(1+\sigma)\left(  1-2\sigma\right)
K_{I}}{E(2\pi r)^{1/2}}\cos(\theta/2)\label{eps0}%
\end{equation}
where $r$ and $\varphi$ are the polar coordinates counted off from the crack
tip and $\sigma$ is the Poisson's ratio. The parameter $\beta$ is expressed in
terms of $\beta_{0}$ (\ref{fch}, \ref{GLK1}) as follows:
\begin{equation}
\beta=\beta_{0}\left\{  1-\frac{2A^{2}}{E\beta_{0}}\frac{(1-2\sigma
)(1+\sigma)}{1-\sigma}\right\} \label{beta}%
\end{equation}
Equation (\ref{EffectiveGLK}) exhaustively describes the order parameter
dynamics within the PZ. It should be mentioned that (\ref{EffectiveGLK}) can
be obtained by the variation procedure (\ref{Variation}) using the dissipation
function (\ref{DissipationFunction}) and the effective free energy:%

\begin{equation}
F_{\text{eff}}=F_{0}+%
{\displaystyle\int}
\left[  \frac{g}{2}\left(  \nabla\eta\right)  ^{2}+\frac{1}{2}\alpha\eta
^{2}+\frac{1}{4}\beta\eta^{4}+A\eta^{2}\varepsilon_{ii}^{(0)}(\mathbf{r}%
)\right]  d\Omega\label{EffectiveFreeEnergy}%
\end{equation}
derived in the Appendix A.

For the sake of completeness let us also mention that in the equilibrium
($\partial\eta/\partial t=0$), homogeneous ($\Delta\eta=0$) state, away from
the tip ($\varepsilon_{ii}^{(0)}=0$) one finds the bulk phase $\eta=0$ also
referred to as the "mother phase" at $T>T_{c}$, while at $T<T_{c}$ the bulk
"daughter phase", $\eta=\pm\left(  -\alpha/\beta\right)  ^{1/2}$ takes place
\cite{LandauStat}.

Equation (\ref{EffectiveGLK}) represents the model A according to the
Hohenberg-Halperin-Ma scheme \cite{HohenbergHalperinMa}. Its solution is
demonstrated in Section III.

\section{Analytical analysis of the equation of motion for the order
parameter}

\subsection{The automodel regime}

Assuming the crack tip propagating with the velocity $V$ along the $Ox$ axis
and passing to the comoving frame, $x^{\prime}=x-Vt$, $y^{\prime}=y$, one
finds the equation of motion (\ref{EffectiveGLK}) in the form:%

\begin{equation}
g\Delta\eta+\kappa V\frac{\partial\eta}{\partial x^{\prime}}-\left[  \alpha\pm
B\frac{\cos(\theta)}{\sqrt{r^{\prime}}}\right]  \eta-\beta\eta^{3}%
=0\label{automodel}%
\end{equation}
where
\begin{equation}
B=2\sqrt{\frac{2}{\pi}}a\left\vert k\right\vert (1+\sigma)K_{I}>0\label{B}%
\end{equation}
and one chooses the sign "$+$", if $A>0$ and "$-$" in the opposite case.
Further, $r^{\prime}=\left(  x^{\prime2}+y^{\prime2}\right)  ^{1/2}$, and the
Laplace operator is defined as $\Delta=\partial^{2}/x^{\prime2}+\partial
^{2}/y^{\prime2}$. From here on we only use the comoving frame and, therefore,
omit the primes.

If one describes a transformational PZ $\eta(\mathbf{r})\neq0$ embedded into
the matrix of the bulk phase ($\alpha>0$, $\eta=0$), localized at the crack
tip, $\left(  x,y\right)  =0$, while vanishing away from it, the boundary
condition takes the form:
\begin{equation}
\eta(\infty)=0\label{BK-LPT}%
\end{equation}
Because its physical origin is related to atomic coordinates which must be
limited, the order parameter is everywhere finite: $\left\vert \eta\right\vert
<\infty$ \cite{Note1}.

For the description of the order parameter distribution embedded into the
matrix of the daughter phase: ($\alpha<0$; $\eta=(-\alpha/\beta)^{1/2}\neq0$),
one needs to use the boundary condition:
\begin{equation}
\eta(\infty)=(-\alpha/\beta)^{1/2}\label{BK-deep}%
\end{equation}

Below to study the problem at hand we employ methods of the bifurcation theory.

\subsection{Bifurcation theory: a brief review}

For the convenience of the reader let us first shortly recite some key results
of theory of bifurcations \cite{Vainberg} which we use in the following
argumentation. Let us consider a nonlinear equation that can be written in the form:%

\begin{equation}
\hat{L}(\alpha)\eta=\hat{N}(\eta)\label{L}%
\end{equation}
where $\hat{L}(\alpha)$ is a linear operator depending upon the parameter
$\alpha$, $\eta=\eta(\mathbf{r})$ is a dependent function and $\hat{N}(\eta)$
is a nonlinear operator, such that $\hat{N}(0)=0$. In general both $\hat
{L}(\alpha)$ and $\hat{N}$ can be differential, integral or
integro-differential operators. In the case considered in our further study
$\hat{L}(\alpha)$ is a differential operator, while $\hat{N}$ is a polynomial.

One can see that equation (\ref{L}) has a trivial solution $\eta=0$. Assume
that it is stable at some $\alpha>0$. The trivial solution of (\ref{L})
becomes unstable, as soon as $\alpha$ reaches $\alpha_{\ast}>0$, equal to the
first eigenvalue, $\alpha_{\ast}=\alpha_{1}$, of the linearized equation
(\ref{L}):%
\begin{equation}
\hat{L}(\alpha_{n})\Psi_{n}(\mathbf{r})=0\label{LEigen}%
\end{equation}
Here $\alpha_{n}$ are the eigenvalues belonging to the discrete spectrum of
the equation (\ref{LEigen}), if any, $\Psi_{n}(\mathbf{r})$ are their
corresponding eigenfunctions and $n=1,2,...$are the natural numbers. In
analogy with quantum mechanics $\alpha_{\ast}\equiv\alpha_{1}$ and $\Psi
_{\ast}(\mathbf{r})\equiv\Psi_{1}(\mathbf{r})$ are referred here to as the
"ground state" eigenvalue and eigenfunction.

In the close vicinity of the bifurcation point one can obtain the
asymptotically-exact, overcritical solution of Eq. (\ref{L}) in the form of a
series in terms of two small parameters: the amplitude, $\xi$, and the
"distance" form the bifurcation point, $\alpha-\alpha_{\ast}$. The bifurcation
theory \cite{Vainberg} ensures, however, that its main term always takes the form:%

\begin{equation}
\eta(\mathbf{r})\ \approx\xi\Psi_{\ast}(\mathbf{r})+O(\xi^{3}%
)\label{LSolution}%
\end{equation}
Here, $\xi$ is the amplitude to be determined from the nonlinear equation
(\ref{L}). This can be done in several ways. Making use of (\ref{LEigen}) and
substituting the main term of the solution (\ref{LSolution}) one can, for
example, represent (\ref{L}) in the form: $\xi\times\lbrack\hat{L}%
(\alpha)-\hat{L}(\alpha_{\ast})]\Psi_{\ast}=\hat{N}(\xi\Psi_{\ast})$.
Multiplying scalarly its both parts by $\Psi_{\ast}$ one finds:%
\begin{equation}
\left\langle \left[  \hat{L}(\alpha)-\hat{L}(\alpha_{\ast})\right]  \Psi
_{\ast},\Psi_{\ast}\right\rangle \xi=\left\langle \hat{N}(\xi\Psi_{\ast}%
),\Psi_{\ast}\right\rangle \label{BranchingMain}%
\end{equation}
where we use the notation: $\left\langle f,g\right\rangle =\int f(\mathbf{r}%
)g(\mathbf{r})d\Omega$ for the scalar product of two functions, $f(\mathbf{r}%
)$ and $g(\mathbf{r})$, in the Hilbert space. (\ref{BranchingMain}) referred
to as a "branching equation" \cite{Vainberg}, represents a nonlinear equation
with respect to $\xi$, only valid if $\xi$ is small. Its solution yields the
amplitude $\xi$, thus, giving simultaneously the solution (\ref{LSolution}) of
the bifurcation problem valid in the close vicinity of the bifurcation point.
Full details, theorems and their proofs one can find in the book of Vainberg
and Trenogin \cite{Vainberg}.

We would like to stress, that though the above recipe heavily involves the
solution of the linear equation (\ref{LEigen}), it represents in fact the
solution of the nonlinear equation (\ref{L}).

\subsection{The process zone embedded into the mother phase}

\subsubsection{ The bifurcation condition}

Let us assume $\alpha>0$ and look for the point of instability of the trivial
solution $\eta=0$ describing the homogeneous mother phase. It can be done
using Eq. (\ref{automodel}) within the approach formulated in the previous Section.

The correspondence between (\ref{L}) and (\ref{automodel}) is established as
follows:
\begin{equation}
\hat{L}(\alpha)\eta=g\Delta\eta+\kappa V\frac{\partial\eta}{\partial
x^{\prime}}-\left[  \alpha\pm B\frac{\cos(\theta)}{\sqrt{r^{\prime}}}\right]
\eta=0\label{LofAlfa}%
\end{equation}%
\begin{equation}
\hat{N}(\eta)=\beta\eta^{3}\label{deltaL}%
\end{equation}

Let us now apply the above recipe of the bifurcation theory \cite{Vainberg} to
the case at hand. First of all one finds that at $k>0$ (corresponding to the
sign "$+$") equation (\ref{LEigen}) has no discrete spectrum meaning that the
solution $\eta=0$ of the equation (\ref{automodel}) is stable at $\alpha>0$.

In contrast to that, at $k<0$ (the sign "$-$") Eq. ( \ref{LEigen}) has a
discrete spectrum at $\alpha>0$ implying that Eq. (\ref{automodel}) exhibits
an instability. Let us consider the latter case.

At large value of $\alpha$ equation (\ref{automodel}) has the trivial solution
$\eta=0$. Below the bifurcation point one finds a non-trivial solution
$\eta=\eta(\mathbf{r})\neq0$ and $\sigma_{ik}(\mathbf{r})=\sigma_{ik}%
^{(0)}(\mathbf{r})+O(\eta^{2})$, where $\eta$ is small. According to its
definition (\ref{fch}) $\alpha$ is expressed in terms of temperature:
$\alpha=a(T-T_{\text{c}})$. Equation (\ref{alfa2}) yields, thus, the
temperature, $T_{\ast}$, of LPT at the crack tip. The exact solution of the
equation (\ref{LEigen}) with $\hat{L}(\alpha)$ given by (\ref{LofAlfa}) is
given in the next Section.

\subsubsection{Exact results for the eigenvalue and eigenfunctions of Eq.
(\ref{LofAlfa})}

Solution of (\ref{LEigen}) plays as we see an outstanding role defining both
the bifurcation point, $\alpha_{\ast}=\alpha_{1}$, and the overcritical
solution, (\ref{LSolution}) of the nonlinear equation (\ref{L}). Let us solve it.

The linear part of equation (\ref{automodel}) takes the form:%

\begin{equation}%
\genfrac{\{}{.}{0pt}{}{g\Delta\Psi_{n}+\kappa V\times\partial\Psi_{n}/\partial
x-\left[  \alpha_{n}-B\frac{\cos(\theta/2)}{r^{1/2}}\right]  \Psi
_{n}=0;}{\text{ }\Psi_{n}(\infty)=0;\text{ }\left\vert \Psi_{n}\right\vert
<\infty}%
\label{auxilliary2}%
\end{equation}

\begin{figure}
[ptb]
\begin{center}
\includegraphics[
height=2.1421in,
width=2.5979in
]%
{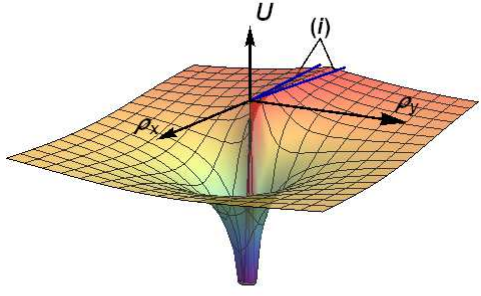}%
\caption{Potential $U(\rho,\theta)=-\cos(\theta/2)/\rho^{1/2}$ of the
Schr\"{o}dinger equation (\ref{Schredinger2}). Here $\rho_{x}=x/R$ and
$\rho_{y}=y/R$. (i) indicates the position of the crack tip.}%
\label{Potential}%
\end{center}
\end{figure}

Making the substitution:%

\begin{equation}
\Psi_{n}(\mathbf{r})=\exp\left(  -\frac{\kappa V}{2g}x\right)  \times\psi
_{n}(\mathbf{r})\label{change}%
\end{equation}
one proceeds to the equation in terms of $\psi_{n}(\mathbf{r})$:%

\begin{equation}
g\Delta\psi_{n}-\left[  \alpha_{n}+\frac{\kappa^{2}V^{2}}{4g}-B\frac
{\cos(\theta/2)}{\sqrt{r}}\right]  \psi_{n}=0\label{Schredinger1}%
\end{equation}
The parameter%

\begin{equation}
R_{1}=\left(  \frac{g}{B}\right)  ^{2/3}=\frac{\pi^{1/3}}{2}\left[  \frac
{g}{a\left\vert k\right\vert (1+\sigma)K_{\text{I}}}\right]  ^{2/3}\label{R1}%
\end{equation}
fixes the characteristic size of the distribution. Passing to dimensionless
cylindrical coordinates $\theta$ and $\mathbf{\rho}=\mathbf{r}/R_{1}$ one
transforms (\ref{Schredinger1}) into the following equation:%

\begin{equation}
\frac{\partial^{2}\psi_{n}}{\partial\rho^{2}}+\frac{1}{\rho}\frac{\partial
\psi_{n}}{\partial\rho}+\frac{1}{\rho^{2}}\frac{\partial^{2}\psi_{n}}%
{\partial\theta^{2}}-\left(  \lambda_{n}-\frac{\cos(\theta/2)}{\sqrt{\rho}%
}\right)  \psi_{n}=0\label{Schredinger2}%
\end{equation}
where $\lambda_{n}$ ($n=1,2...$) represents the eigenvalues of
(\ref{Schredinger2}). They are related to $\alpha_{n}$ as follows:
\begin{equation}
\lambda_{n}=\frac{g^{1/3}}{B^{4/3}}\left(  \alpha_{n}+\frac{\kappa^{2}V^{2}%
}{4g}\right) \label{Lambda1}%
\end{equation}
Eq. (\ref{Schredinger2}) represents a 2D Schr\"{o}dinger equation with an
anisotropic potential $U(\rho,\theta)=-\cos(\theta/2)/\sqrt{\rho}$ shown in
Fig. \ref{Potential}.%

Let us look for the solution of Eq. (\ref{Schredinger2}) in the form
$\Psi(\rho,\theta)=\exp(-\varepsilon z)f(z)$, where $z=\rho^{1/2}\cos
(\theta/2)$. This yields the equation:%
\[
f^{\prime\prime}(z)-4\varepsilon zf^{\prime}(z)-4(\varepsilon-z)f(z)=0
\]
Making use of the transformation $f=\varphi(\zeta)\exp(z/\varepsilon)$ with
$\zeta=(z-1/4\varepsilon^{2})(\varepsilon/2)^{1/2}$ brings one to the equation
in terms of $\varphi$:%

\begin{equation}
\varphi^{\prime\prime}(\zeta)-2\zeta\varphi^{\prime}(\zeta)+m\varphi
(\zeta)=0;\text{ \ \ }m=2[1-(4\varepsilon^{3})^{-1}]\label{EquationFi}%
\end{equation}
The latter represents a Hermitian equation with the eigenvalue $m$ only taking
non-negative, integer, even values: $m=0,2,4...$ Using (\ref{EquationFi}) one
finds that the condition $\varepsilon>0$ only fulfills for the ground state
solution $m=0$ yielding $\varepsilon=2^{-2/3}$ and $\varphi(\zeta)=const$. Now
one can return to the initial variables and write down the solution of the
equation (\ref{Schredinger2}) corresponding to $m=0$:%

\begin{equation}
\lambda_{1}=\frac{1}{2\sqrt[3]{2}}\approx0.397\text{; \ \ \ }\psi
_{1}(\mathbf{\rho})=\exp\left\{  -\frac{\rho}{\sqrt[3]{4}}+\sqrt[3]{4}%
\sqrt{\rho}\cos(\theta/2)\right\} \label{Lambda2}%
\end{equation}
The latter is shown in Fig. \ref{Distribution} (A). Using (\ref{Lambda1},
\ref{Lambda2}) one finds the ground state eigenvalue, $\alpha_{\ast}$:%

\begin{equation}
\alpha_{\ast}=\frac{1}{2\sqrt[3]{2}}\frac{B^{4/3}}{g^{1/3}}-\frac{\kappa
^{2}V^{2}}{4g}\label{alfa2}%
\end{equation}
yielding finally the ground eigenvalue (\ref{alfaSt}) and the eigenfunction
(\ref{psiStar}).

\subsection{The process zone embedded into the mother phase}

\subsubsection{Bifurcation point}

The above solution is valid at $k<0$. It gives the ground state eigenvalue:
\begin{equation}
\alpha_{\ast1}=a(T_{\ast1}-T_{c})=\frac{2^{2/3}}{\pi^{2/3}g^{1/3}}\left[
a\left\vert k\right\vert (1+\sigma)K_{I}\right]  ^{4/3}-\frac{\kappa^{2}V^{2}%
}{4g}\text{\ \ }\label{alfaSt}%
\end{equation}
and the eigenfunction:
\begin{equation}
\Psi_{\ast1}(\mathbf{r})=\exp\left[  -\frac{\nu_{1}}{\sqrt[3]{4}}\frac
{r}{R_{1}}\cos(\theta)-\frac{r}{\sqrt[3]{4}R_{1}}+\sqrt[3]{4}\sqrt{\frac
{r}{R_{1}}}\cos(\theta/2)\right] \label{psiStar}%
\end{equation}
where $\nu_{1}=V/V_{c1}$ is the dimensionless velocity, and the critical
velocity $V_{c1}$ is defined as
\begin{equation}
\text{ \ \ \ }V_{c1}(K_{I})=\frac{2^{4/3}g^{1/3}}{\pi^{1/3}\kappa}\left[
a\left\vert k\right\vert (1+\sigma)K_{I}\right]  ^{2/3}%
\label{CriticalVelocity}%
\end{equation}
Its physical sense will be discussed below.

Let us note that Eq. (\ref{alfaSt}) yields the case $T_{\ast1}-T_{c}>0$ shown
in Fig. \ref{fig1} (B).

Assigning the exponent to $-1$ one finds the size, $L_{f}$, of the order
parameter distribution in front of the tip:%
\begin{equation}
L_{f1}=2^{2/3}\frac{(3+\nu_{1})+2\times(2+\nu_{1})^{1/2}}{(1+\nu_{1})^{2}%
}R_{1}\label{Lf}%
\end{equation}

\begin{figure}
[ptb]
\begin{center}
\includegraphics[
height=1.8654in,
width=2.7017in
]%
{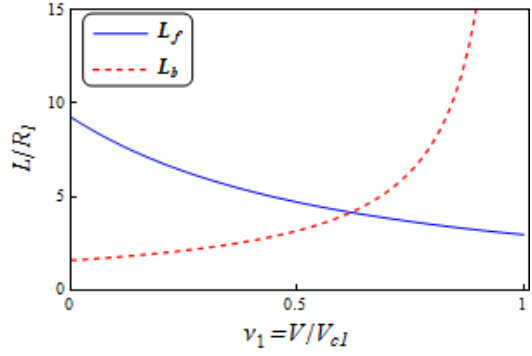}%
\caption{Dependence of $L_{f}$ (solid line) and $L_{b}$ (dashed line) upon
$\nu_{1}$.}%
\label{fig2}%
\end{center}
\end{figure}

At the tip of the motionless crack ($\nu_{1}=0$) one finds the order parameter
distribution size $L_{f1}\approx9.25R_{1}$ decreasing down to $L_{f1}%
\approx2.96R_{1}$ at $\nu_{1}=1$. The dependence $L_{f1}=L_{f1}(\nu_{1})$ is
shown in Fig. \ref{fig2}.%

\subsubsection{The overcritical solution}

Substituting the obtained eigenfunction (\ref{psiStar}) as well as
(\ref{LofAlfa}) and (\ref{deltaL}) into the branching equation
(\ref{BranchingMain}) one obtains%

\begin{equation}
I_{2}\left(  \alpha-\alpha_{\ast1}\right)  \xi+I_{4}\beta\xi^{3}%
=0\label{branching2}%
\end{equation}
where the factors $I_{n}$ ($n=2$, $4$) are the integrals over the whole plane:%

\begin{equation}
I_{n}(\nu_{1})=%
{\displaystyle\int}
\Psi_{\ast}^{n}(\mathbf{\rho,}\nu_{1})d^{2}\rho\label{In}%
\end{equation}
depending upon $\nu_{1}$. 

It should be noted that substitution of (\ref{LSolution}, \ref{psiStar}) into
the effective free energy, (\ref{EffectiveFreeEnergy}) yields the LPT free
energy:%
\begin{equation}
F_{\text{eff}}=F_{0}+R_{1}^{2}\left[  \frac{I_{2}\left(  \alpha-\alpha_{\ast
1}\right)  }{2}\xi^{2}+\frac{I_{4}\beta}{4}\xi^{4}\right] \label{Feff}%
\end{equation}
As expected, its minimization with respect to $\xi$ brings one back to the
branching equation (\ref{branching2}) giving the alternative way to build the
overcritical solution.

The solution of the branching equation has the form:
\begin{equation}
\xi=%
\genfrac{\{}{.}{0pt}{}{0,\text{
\ \ \ \ \ \ \ \ \ \ \ \ \ \ \ \ \ \ \ \ \ \ \ \ \ \ \ \ \ \ \ \ \ \ }%
\alpha>\alpha_{\ast1}}{\pm\left(  \frac{I_{2}}{I_{4}}\right)  ^{1/2}\left(
\frac{\alpha_{\ast1}-\alpha}{\beta}\right)  ^{1/2},\text{ \ \ }\alpha
\leq\alpha_{\ast1}}%
\label{ksiSecondOrder}%
\end{equation}
The integrals $I_{n}(\nu_{1})$ cannot be obtained analytically at $\nu_{1}%
\neq0$. We calculated their ratio numerically by using a standard NIntegrate
routine of Mathematica 10.1 \cite{Wolfram} employing the even-odd subdivision
method with the local adaptive strategy.%

\begin{figure}
[ptb]
\begin{center}
\includegraphics[
height=1.4806in,
width=2.2857in
]%
{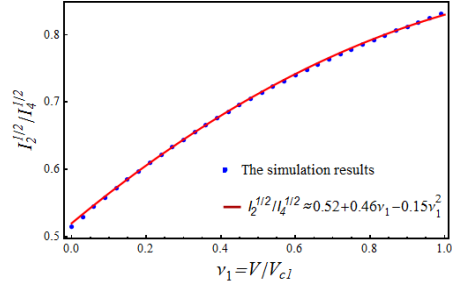}%
\caption{The ratio $I_{2}^{1/2}/I_{4}^{1/2}$ obtained numerically as the
function of the dimensionless velocity $\nu$ and its fitting by the polynomial
(\ref{Appr}).}%
\label{figI2I4}%
\end{center}
\end{figure}

The ratio $I_{2}^{1/2}/I_{4}^{1/2}$ obtained this way is shown in Fig.
\ref{figI2I4} versus the dimensionless velocity $\nu_{1}$. The numerical
result can be accurately fitted by a simple polynomial:%
\begin{equation}
\left(  \frac{I_{2}}{I_{4}}\right)  ^{1/2}\approx0.52+0.46\nu_{1}-0.15\nu
_{1}^{2}\label{Appr}%
\end{equation}
Thus, the asymptotically exact overcritical solution (\ref{LSolution}) is obtained.%

\begin{figure}
[ptb]
\begin{center}
\includegraphics[
height=2.0038in,
width=4.4322in
]%
{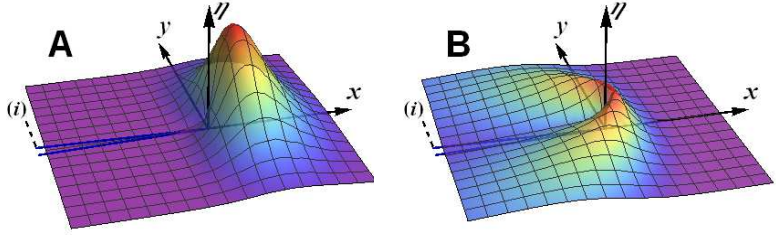}%
\caption{Spatial distribution of the order parameter, $\eta(x,y)$, in the
vicinity of the motionless (A) and propagating (B) crack. Two blue lines (i)
indicate the crack tip position.}%
\label{Distribution}%
\end{center}
\end{figure}

\subsubsection{The critical velocity}

The result (\ref{ksiSecondOrder}) implies that the LPT only takes place at
$\alpha<\alpha_{\ast}$, while at $\alpha>$ $\alpha_{\ast}$ the crack tip is
undressed. Since at $\alpha<0$ the whole bulk of the solid transforms into the
phase $\eta\neq0$, the domain in which LPT takes place is restricted to the
interval: $0<\alpha\leq\alpha_{\ast}$. The latter result has a very important
consequence. Making use of Eq. (\ref{alfa2}) one finds that at, $V=V_{c1}$ the
LPT disappears. In other words the LPT can only exist at the crack tip at
$V<V_{c1}$, while at $V\geq V_{c1}$ it vanishes. This property is fundamental
for any LPT, both of the second and of the first order. It is a direct
consequence of the fact that the order parameter has its own dynamics
exhibiting an intrinsic characteristic time, and that as soon as $V\geq
V_{\text{c}1}$ the order parameter in front of the crack tip has no time to
evolve from $\eta=0$ to $\eta\approx\xi$.

The spatial distribution of the order parameter (\ref{LSolution}) is shown in
Fig. \ref{Distribution}. Here (A) shows the order parameter in the vicinity of
the motionless crack tip, while (B) displays that in the case of a propagating
crack. This image is obtained with the velocity value $V=0.5V_{c1}$.

One finds that the order parameter is highly localized in the vicinity
$r\precsim10R_{1}$ of the tip of the motionless crack. In the case of the
moving crack tip (Fig. \ref{Distribution} B) the order parameter distribution
is compressed in front of the tip and stretched in its back with respect to
that at the motionless one. The length, $L_{b}$, of the order parameter
distribution behind the tip of the propagating crack takes the form:
\begin{equation}
L_{b}=\frac{\sqrt[3]{4}R_{1}}{1-V/V_{c1}}\label{Trail}%
\end{equation}
Behind the crack the length, $L_{b}$, diverges, if $V\rightarrow V_{c1}$ (Fig.
\ref{fig2}). It should be noted that at $V\rightarrow V_{c1}$ the amplitude,
$\xi$, vanishes.

Let us summarize the results of the present Section. We have shown that in a
solid with a crack the trivial solution $\eta\equiv0$ is stable at high
temperatures ($T>T_{c}$), but loses its stability at the point $\alpha
=\alpha_{\ast1}>0$ corresponding to a temperature $T_{\ast1}$ somewhat higher
than that of the bulk phase transition: $T_{\ast1}>T_{c}$. This solution
describes a region of the phase $\eta\neq0$ embedded into the matrix
$\eta\equiv0$ representing the transformational PZ at the crack tip. In terms
of temperature its existence is limited to the domain $T_{c}<T<T_{\ast1}$,
while in terms of velocity to $0\leq V\leq V_{c1}$. The transformational PZ at
the tip of the propagating crack is deformed with respect to that of the
motionless crack: the order parameter distribution is compressed in its front,
while stretched in its back.

\subsection{The process zone embedded into the daughter phase}

Let us first find a solution for $\eta(\mathbf{r})$ in the low-temperature
phase ($\alpha<0$). We will assume in addition that $\left\vert \alpha
\right\vert $ is large enough, so that where this solution is stable.
Observing that at $r\gg(g/\alpha)^{1/2}$ the terms $\sim\Delta\eta$ and
$\sim\partial\eta/\partial x$ in (\ref{automodel}) are much smaller than the
others and neglecting them, one finds the approximate distribution $\eta
=\eta_{0}(\mathbf{r})$ in the daughter phase:
\begin{equation}
\eta_{0}(r,\theta)\approx\frac{1}{\beta^{1/2}}\left(  -\alpha-B\frac
{\cos(\theta/2)}{\sqrt{r}}\right)  ^{1/2}\label{LowTemp}%
\end{equation}
To be specific, from two solutions of the equation of state we have chosen a
positive one. Let us now look for the solution perturbation in the form:
$\eta(\mathbf{r})=\eta_{0}(\mathbf{r})+\delta\eta(\mathbf{r})$. Substituting
it into (\ref{automodel}) one finds that the term $\delta\eta(\mathbf{r})$ is
subjected to the equation (\ref{L}) with
\begin{equation}
\hat{L}(\alpha)\delta\eta=g\Delta\delta\eta+\kappa V\frac{\partial\delta\eta
}{\partial x}-\left[  2\left\vert \alpha\right\vert \mp B\frac{\cos(\theta
)}{\sqrt{r}}\right]  \delta\eta\label{LofDeltaEta}%
\end{equation}
and%
\begin{equation}
\hat{N}(\delta\eta)=3\left(  -\alpha\beta-\frac{B\beta\cos(\theta/2)}{\sqrt
{r}}\right)  ^{1/2}\delta\eta^{2}+\beta\delta\eta^{3}\label{NofDeltaEta}%
\end{equation}

In this case at $k<0$ (the sign "$+$" in \ref{LofDeltaEta}) equation
(\ref{LofDeltaEta}) appears to have no discrete spectrum implying that the
solution (\ref{LowTemp}) is stable. The discrete spectrum indicating the
instability, however, exists at $k>0$ (the sign "$-$" in \ref{LofDeltaEta}).
Below we consider this latter case.

Applying the analysis already described above to the present case one finds
the eigenfunction (\ref{psiStar}), in which instead of $R_{1}$ one should take
a characteristic size $R_{2}$ expressed as:%
\begin{equation}
R_{2}=\left(  \frac{g}{2B}\right)  ^{2/3}=\frac{\pi^{1/3}}{2^{5/3}}\left[
\frac{g}{akK_{\text{I}}(1+\sigma)}\right]  ^{2/3}\label{R2}%
\end{equation}
which is smaller than (\ref{R1}) by the factor $2^{-2/3}$. The bifurcation
point has the form:%
\begin{equation}
\alpha_{\ast2}=-\frac{1}{2^{1/3}\pi^{2/3}g^{1/3}}\left[  ak(1+\sigma
)K_{I}\right]  ^{4/3}+\frac{\kappa^{2}V^{2}}{8g}\label{LowTempBifr}%
\end{equation}
such that a non-trivial solution $\delta\eta(\mathbf{r})=\xi_{2}\Psi_{\ast
}(\mathbf{r})$ takes place at $0\geq\alpha\geq\alpha_{\ast2}$ corresponding to
the phase diagram shown in Fig. \ref{fig1} (A). Its amplitude, $\xi_{2}$,
should be determined form the branching equation. Analogously to the previous
case the relation (\ref{LowTempBifr}) gives rise to the critical velocity:%
\begin{equation}
V_{c2}=\frac{2^{4/3}g^{1/3}}{\pi^{1/3}\kappa}\left[  akK_{I}(1+\sigma)\right]
^{2/3}\label{Vcr2}%
\end{equation}
limiting the existence of the high-temperature PZ in the matrix of the
low-temperature phase. The latter is by the factor of $2^{2/3}$ larger than
$V_{\text{c}1}$.

Since the overcritical solution (\ref{LSolution}) for $\delta\eta$ is
determined by the same eignfunction (\ref{psiStar}) as that for the
high-temperature case, the distribution, $\delta\eta(\mathbf{r})=\xi_{2}%
\Psi_{\ast}(\mathbf{r})$ has the same form as that of the order parameter in
the high-temperature PZ. The distribution is shown in Fig. (\ref{Distribution}%
). Analogously to the high-temperature case the distribution, $\delta
\eta(\mathbf{r})$, at the tip of the propagating crack is compressed in its
front, stretched backwards and the relation (\ref{Trail}) holds. The
amplitude, $\xi_{2}$, should be determined from the branching equation.

These further results are easy, but rather cumbersome to derive, and we,
therefore, give the calculations in Appendix C.

Let us summarize the findings of this Section. We found that the inhomogeneous
solution $\eta_{0}(\mathbf{r})$ describing the order parameter distribution in
the low-temperature phase is stable at $\alpha<\alpha_{\ast2}<0$. At
$\alpha=\alpha_{\ast2}<0$ this solution becomes unstable and the solution,
$\eta_{0}(\mathbf{r})+\delta\eta(\mathbf{r})$, branches decreases off from
$\eta_{0}$, the signs of $\eta_{0}$ and $\delta\eta$ being different. This
means that the zone with the high-temperature phase, $\eta\equiv0$, emerges at
the crack tip at $\alpha=\alpha_{\ast2}$ and exists within the domain
$0\leq\alpha\leq\alpha_{\ast2}$ and at $0\leq V\leq V_{\text{c2}}$.

\section{Simulation}

\subsection{Rescaling}

Based on the bifurcation theory our analytical results are only valid in the
close vicinity of the bifurcation point. To study the problem far from the
bifurcation we simulated solution of the equation (\ref{automodel}). We report
below the simulation of the daughter phase PZ embedded into the matrix of the
mother phase.

The equation (\ref{automodel}) has been rescaled making the variables
dimensionless and minimizing the control parameters number: $x\rightarrow
d_{1}x_{1}$, \ $y\rightarrow d_{1}y_{1}$ and \ \ $\eta(x,y)\rightarrow
d_{2}u(x_{1},y_{1})$ with the scaling factors $d_{1}=R_{1}$ and $d_{2}%
=B^{2/3}/g^{1/3}\beta^{1/2}$. The rescaled equation (\ref{automodel}) takes
the form:
\begin{equation}
\Delta_{1}u+2^{1/3}\nu\frac{\partial u}{\partial x_{1}}-\left[  q-\frac
{\cos(\theta/2)}{r_{1}^{1/2}}\right]  u-u^{3}=0\label{Sim1}%
\end{equation}
where $\Delta_{1}=\partial^{2}/\partial x_{1}^{2}+\partial^{2}/\partial
y_{1}^{2}$, $r_{1}=(x_{1}^{2}+y_{1}^{2})^{1/2}$, $d\Omega_{1}=dx_{1}dy_{1}$
and the dimensionless parameter $q$ is expressed as follows:%

\[
q=\frac{g^{1/3}}{B^{4/3}}\alpha
\]
\qquad Let us note that in terms of the rescaled parameters $(q,\nu)$ the
analytical expression for the bifurcation condition (\ref{alfaSt}) takes the
form
\begin{equation}
q_{c}=\frac{1-\nu^{2}}{2\times2^{1/3}}\label{AcOfNu}%
\end{equation}
while the overcritical solution (\ref{psiStar}, \ref{ksiSecondOrder}) at
$q<q_{c}$ is expressed as:%
\begin{equation}
u(r_{1},\theta)=\left[  \frac{I_{2}\left(  q_{c}-q\right)  }{I_{4}}\right]
^{1/2}\exp\left\{  -4^{-1/3}r_{1}\left[  1-\nu\cos(\theta)\right]
+4^{1/3}r_{1}^{1/2}\cos(\theta/2)\right\} \label{OPRescaled}%
\end{equation}

\subsection{Results}

We used a pseudo-time stepping approach representing a version of the
iteration method. Its description along with software technical details and
settings are given in the Appendix C. The pseudo-time stepping approach
converged away from the points: $q=0$ and $q=q_{c}$ (\ref{AcOfNu}). In
practice, we obtained a good convergence at $q>0.1$, above the points of the
global bifurcation. Closer to the point $q=0$ we were unable to get an
equilibrium solution. In the vicinity to the point $q=q_{c}$ the method
produced a small regular error discussed in details below. Apart from that the
method exhibited a good convergence, enabling us to study the distribution of
the rescaled order parameter, $u(x_{1},y_{1})$ in the vicinity of the crack tip.

\begin{figure}
[ptb]
\begin{center}
\includegraphics[
height=2.6403in,
width=4.1909in
]%
{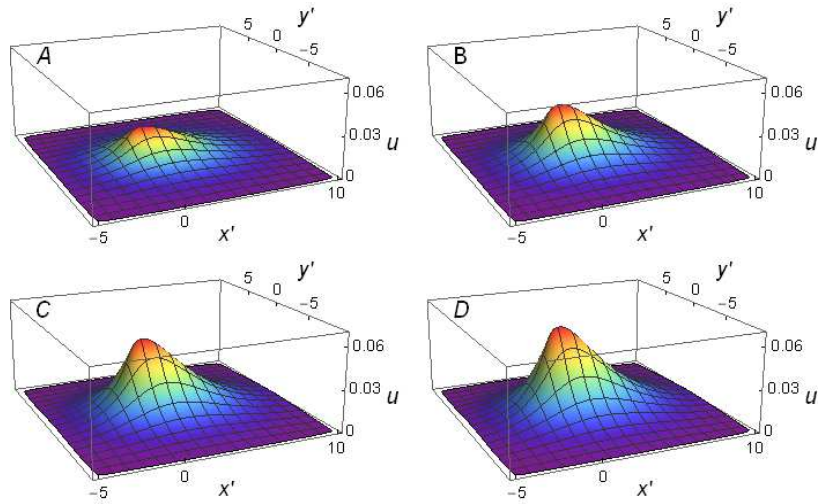}%
\caption{Order parameter distribution obtained by simulations at $\nu=0$ and
various values of $q_{c}-q$: (A) $1.5\times10^{-4}$, (B) $6.5\times10^{-4}$,
(C) $12\times10^{-4}$ and (D) $16\times10^{-4}$.}%
\label{op3D}%
\end{center}
\end{figure}

Figure \ref{op3D} shows the distributions of the rescaled order parameter at
the tip of a motionless crack ($\nu=0$) at the successively increasing values
of the difference $q_{c}-q>0$ below the bifurcation point. As expected the
order parameter grows with $q_{c}-q$, its distribution being close to that
described by Eq. (\ref{OPRescaled}) at $\nu=0$ (shown in Fig.
\ref{Distribution} A).%

\begin{figure}
[ptb]
\begin{center}
\includegraphics[
height=2.2632in,
width=3.4229in
]%
{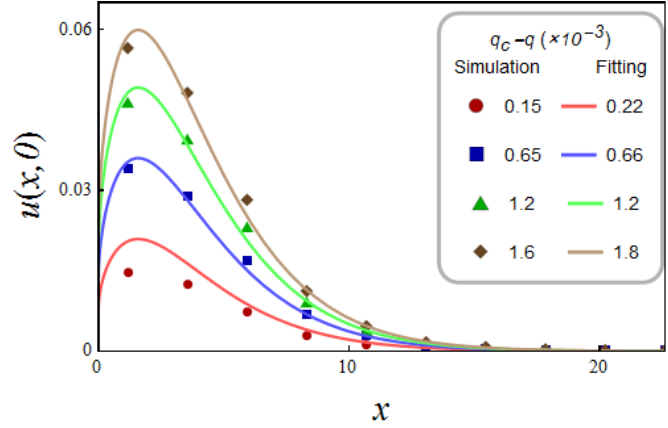}%
\caption{Cross-section of the order parameter distribution along the plane
$y=0$ shown at $x>0$ corresponding to $\nu=0$ and various values of $q$. Dots
show the simulation results at different values of $q_{c}-q$, while the solid
line display their fittings with the exptession (\ref{OPRescaled}). The legend
for the dots indicate the $q$ values used in the simulation, while that for
the solid lines yields $q$ values obtained by the fitting.}%
\label{comparison}%
\end{center}
\end{figure}

To check the proximity of the simulated results to the analytical one we
plotted the cross-sections of the above solutions by the plane $(x_{1},u)$
(Fig. \ref{comparison}). This has been done by sampling the solution points
from the layer with the thickness $1.5$ along the plane $y_{1}=0$. Because the
simulation results exhibit a regular shift of the bifurcation point we plot
the order parameter versus the distance from the bifurcation, $q_{c}-q$,
rather then $q$. The solid lines in Fig. \ref{comparison} show the
cross-section of the analytical solution (\ref{OPRescaled}) at the same values
of $q_{c}-q$. One can see that close to the bifurcation point the solution
obtained by the simulation is rather close to the analytical one (Fig.
\ref{comparison}).%

\begin{figure}
[ptb]
\begin{center}
\includegraphics[
height=3.5743in,
width=5.0678in
]%
{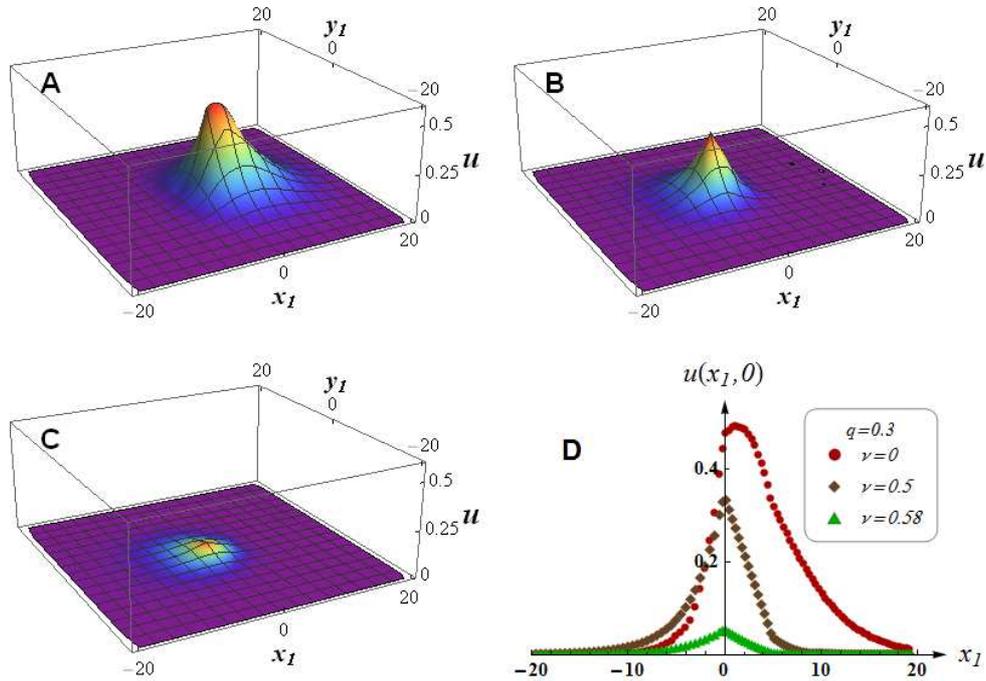}%
\caption{Order parameter distribution in the vicinity of the tip of the crack
propagating in the positive direction of $x$ along the line $y=0$. All images
have been obtained by simupation with $q=0.3$. (A) shows the motionless case
$\nu=0$, while (B) and (C) display the case of the propagating crack:
$\nu=0.5$ (B) and $\nu=0.58$ (C). The image (D) shows the cross-secion of the
distributions shown in (A-C) by the plane $y=0$.}%
\label{movingCrack}%
\end{center}
\end{figure}

The distribution of the rescaled order parameter, $u(x_{1},y_{1})$ at the tip
of the propagating crack ($\nu_{1}\geq0$) at the same value of $q$ is shown in
Fig. \ref{movingCrack}. One can see that the crack motion transforms the
distribution in several ways. First, with the increasing of the dimensionless
velocity, $\nu_{1}=V/V_{c1}$, it becomes lower and vanishes as soon as the
velocity achieves the bifurcation line. Besides, the order parameter
distribution is compressed in front of and stretched out behind the tip with
respect to that at the tip of the motionless crack.%

\begin{figure}
[ptb]
\begin{center}
\includegraphics[
height=2.207in,
width=3.269in
]%
{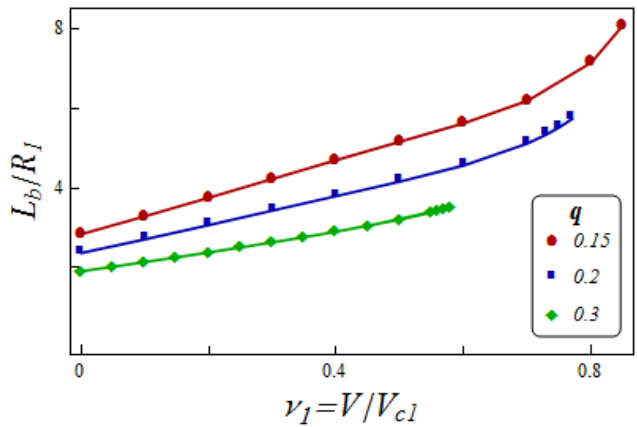}%
\caption{Dependence of the dimensionless length, $L_{b}/R_{1}$, of the order
parameter trail upon the dimensionless velocity, $\nu_{1}=V/V_{c1}$ for
several values of the parameter $q$. For each value the data has been taken
unltil the bifurcation line has been approached. Dots show the simulation
result, while the solid line is guiding the eye.}%
\label{trailLength}%
\end{center}
\end{figure}

The latter is in line with the predictions of the analytical solution
(\ref{OPRescaled}) at $x_{1}<0$, $y_{1}=0$:
\begin{equation}
u(x_{1},0)\sim\exp\left[  -4^{-1/3}x_{1}\left(  1-\nu\right)  \right]  \text{;
\ \ }x_{1}<0\label{tail}%
\end{equation}
describing the distribution behind the crack tip. The distribution length
diverges as $4^{1/3}/(1-\nu)$ as $\nu\rightarrow1$ (see also \ref{Trail}). The
simulation, indeed, looks close to the exponents (Fig. \ref{movingCrack} D).
To check this we fitted the backward part of the distributions by the
exponents $u(x_{1}<0,0)\sim\exp(x_{1}R_{1}/L)$. Figure \ref{trailLength}
displays the dependence of $L/R_{1}$ upon $\nu_{1}$ at different values of the
control parameter $q$ obtained by the simulations.%

It should be noted that the trail length, indeed, increases with the velocity,
but does not follow the behavior (\ref{Trail}) predicted by our analytical
approach. This is because the analytical prediction is only valid very close
to the bifurcation point, which is not the case for most of the points shown
in Fig. \ref{trailLength}.

The plane $(\nu_{1},q)$ is divided into two regions. In the region I
($q<q_{c}$) the transformational PZ at the crack tip exists, while it vanishes
at the boundary ($q=q_{c}$) and does not exist in the region II ($q\geq q_{c}
$). To check this we made simulations by fixing the value of $\nu_{1}$ and
evaluating the maximal order parameter value, $u_{\max}$, at different $q$
points. One can see that at large $q$ values $u_{\max}$ vanishes, while
emerging after a certain threshold.%

\begin{figure}
[ptb]
\begin{center}
\includegraphics[
height=2.386in,
width=3.5449in
]%
{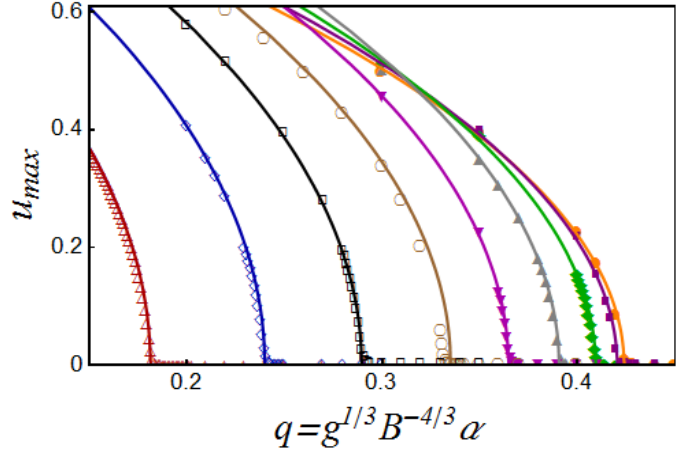}%
\caption{Dependence of the maximm value of the rescaled order parameter,
$u_{\max}$, upon the control parameter $q$ at different values of the
dimensionless velocity $\nu_{1}$: disks show $\nu_{1}=0$, squares - $0.1$,
filled diamonds - $0.2$, filled vertex-up triangles - $0.3$, vertex-down
triangles - $0.4$, open circles - $0.5$, open squares - $0.6$, open diamonds -
$0.7$, open vertex-up triangles - $0.8$. The solid lines show their fitting to
the function (\ref{fitting}).}%
\label{UofA}%
\end{center}
\end{figure}

\begin{figure}
[ptb]
\begin{center}
\includegraphics[
height=2.4223in,
width=3.7196in
]%
{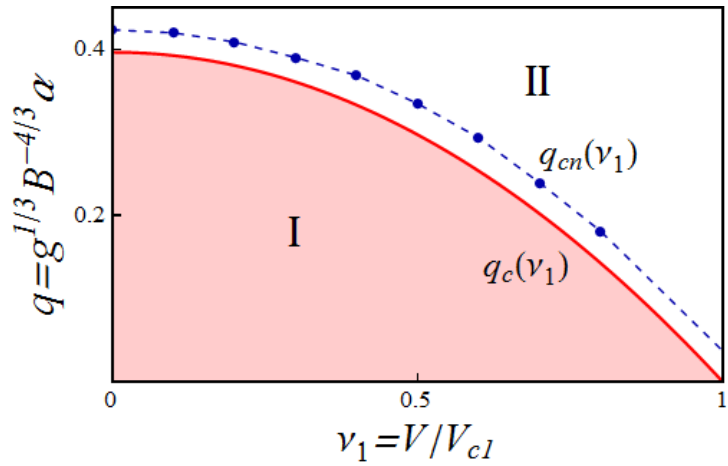}%
\caption{Dynamic phase diagram divides the $(\nu_{1},q)$ plane into the part I
in which the transformational process zone is present at the crack tip and II
without the zone. The dots show the results of simulation. They are connected
by a dahsed line to guide the eye. The solid line shows the theoretical
boundary given by the expression (\ref{AcOfNu}).}%
\label{AcOfNuPlot}%
\end{center}
\end{figure}

The values of these threshold have been extracted from the above data by
fitting to the function:
\begin{equation}
u_{\max}(q)=%
\genfrac{\{}{.}{0pt}{}{0\text{, \ \ \ \ \ \ \ \ \ \ \ \ \ \ \ \ \ \ \ \ }%
q>q_{cn}}{u_{0}\left(  q_{cn}-q\right)  ^{1/2}\text{, \ }q\leq q_{cn}}%
\label{fitting}%
\end{equation}
where $u_{0}$ and $q_{cn}$ are the fitting parameters, and $q$ is the
variable. The parameter $q_{cn}$ represents, therefore, the bifurcation value
obtained from the simulations. The subscript "n" stays for "numeric", to
distinguish $q_{cn}$ from the analytically obtained bifurcation boundary,
$q_{c}$. The dynamic phase diagram obtained this way is shown in Fig.
(\ref{AcOfNuPlot}).%

One can see that the simulation gives the values of the boundary points that
are regularly shifted over about $5\%$ upwards with respect to the line
obtained analytically. We did not succeed to obtain values close to the line
$q_{c}(\nu)$. This is due to the singularity $\sim r^{-1/2}$ in the equation
(\ref{Sim1}) \cite{CracksSimulations}.\qquad

\section{Discussion}

\subsection{Our findings}

We formulated a natural approach to describe the LPT at the tip of the crack
taking into account the order parameter, the internal degree of freedom
responsible for the phase transition. The LPT takes place as the response of
the solid to the local stress inhomogeneity. This suggests that any type of
stress concentrator may generate LPT. In addition to the crack tips the role
of such stress concentrators play surfaces, inclusions, grain and twin
boundaries, dislocations and disclinations. It is, further, clear that for any
of this type of the concentrators the LPT (i) decreases the total concentrator
energy and (ii) introduces its dissipation in the process of motion, thus,
influencing its static \cite{Korzhenevskii}, \cite{Boulbitch and Pumpjan} and
dynamic \cite{Our 1} properties.

Among the LPT generating stress concentrators cracks are the most powerful
ones. LPT-induced variation of the cracks properties may considerably alter
the ability of material to resist fracture. This paper focuses on the LPT at
the crack tips. The conclusions of the present paper are based on our
following fundamental findings.

First, we have found the exact solution for the point of bifurcation both in
the case of the motionless and propagating crack.

Second, we established the existence of the critical crack velocity
controlling the disappearance of the LPT zone at high speeds.

Third, we have found an analytical, asymptotically exact solution of the
nonlinear equation describing the order parameter distribution close to the
bifurcation point.

Fourth, by numeric simulation we obtained the solutions away from the
bifurcation point which is in line with our analytic solution.

It should be mentioned here that the present results will be further used for
analysis of the fast crack propagation published separately.

\subsection{Effect on the crack behavior of the second versus the first order
transitions}

In this paper we only addressed the phase transitions of the second order.
Second order phase transitions are generally considered to be soft. This
opinion gives rise to the illusion that they should only have a negligible
effect on the crack dynamics. This point of view is erroneous. The softness of
the second order phase transitions only means that the solid continuously
passes from the state $\eta=0$ to the state $\eta\neq0$ in the transition or
bifurcation point. This is in contrast to the first order transitions where in
the transition point the order parameter exhibits a jump. The effect of the
LPT on the crack dynamics is, however, not related to the order parameter
behavior in the transition point. It depends upon absolute values the order
parameter can achieve within the LPT zone. In the discussion below we will
argue that the zone is typically very wide. This implies that the PZ order
parameter is most often in the saturated state. In this respect the second
order phase transition exhibits no qualitative difference from that of the
first order. In the case of structural phase transitions the order parameter
can be constructed using the atoms displacements from their positions in the
mother phase. Their saturation values may achieve the values smaller, but
comparable to the crystal lattice cell dimensions both in the case of the
first and second order phase transition.

It should be, further, noted that for one thing the value of the bifurcation
point, $T_{\ast}$, is obtained by the analysis of the linear part,
(\ref{LEigen}), of the nonlinear equation for the order parameter. For the
other, the answer to the question, whether the PT is of the second, or first
order depends only upon the structure of the nonlinear terms of the free
energy (\ref{fch}). This suggests that the result (\ref{DeltaTMain}) for
$T_{\ast}$ is valid both for the cases of the first and second order
transitions. Here we take this assumption as granted. A rigorous proof of this
statement will be published elsewhere.

\subsection{Estimates}

\subsubsection{Material parameters and results for selected materials}

The temperature shift value, $\Delta T_{\ast}$, is of a great importance,
since it determines the phase diagram region where the PZ influences the crack
behavior. It is interesting, therefore, to have numerical values of $\Delta
T_{\ast}$ at least for some materials. As the examples let us consider
ferroelectrics BaTiO$_{3}$, PbTiO$_{3}$ and LiNbO$_{3}$.

All the three materials belong to a large family of perovskites, LiNbO$_{3}$
exhibiting a distorted perovskite structure \cite{Lehnert}.

The high-temperature phases of PbTiO$_{3}$ and BaTiO$_{3}$ exhibit a cubic
symmetry. The cubic phase in PbTiO$_{3}$ exists above $850%
\operatorname{K}%
$, while transforming into the tetragonal one at lower temperatures
\cite{Jabarov}. In contrast to that BaTiO$_{3}$ exhibits the cubic phase (II),
tetragonal (III), rhomboherdic (IV) and orthorhombic phase (V). The numeration
of the phases is given according to \cite{Tonkov}. The phase diagram of
BaTiO$_{3}$ can be found in the paper \cite{Ishidate}. All the transitions of
BaTiO$_{3}$ and PbTiO$_{3}$ are of the first order.

The Landau theory enables one to describe possible transitions in the both
materials as generated by the symmetry lowering of the cubic mother phase, the
order parameter being a 3D vector, $(\eta_{1},\eta_{2},\eta_{3})$ associated
with the polarization. In the cubic phase $\mathbf{\eta}=(0,0,0)$. In all the
low-temperature phases showing up in the phase diagrams the order parameter
has only one independent component: $(\eta,0,0)$ in the tetragonal,
$(\eta,\eta,\eta)$ in the rhombohedric, while the orthorhombic phase is
described by $(\eta,\eta,0)$ \cite{Gufan}. Effectively, therefore, the problem
is reduced to a single order parameter enabling one to directly apply the
approach developed in the present paper. It should be noted that the free
energy of BaTiO$_{3}$ and PbTiO$_{3}$ has striction terms giving rise to
nondeviatoric spontaneous strain in addition to the deviatoric one. For the
order of magnitude estimate done below this difference is, however, irrelevant.

LiNbO$_{3}$ possesses the symmetry $R\overline{3}c$ in the high-temperature
paraelectric phase and exhibits a 2nd order transition into the ferroelectric
$R3c$ phase at $1460%
\operatorname{K}%
$ \cite{Lehnert}, the transition being described by a one-component order parameter.

Material parameters of BaTiO$_{3}$, PbTiO$_{3}$ and LiNbO$_{3}$ are summarized
in the Table 1.


\begin{table}
\caption{Material constants of BaTiO$_{3}$, PbTiO$_{3}$ and LiNbO$_{3}$}
\begin{tabular}
[c]{|c|c|c|c|}\hline
& BaTiO$_{3}$ & PbTiO$_{3}$ & LiNbO$_{3}$\\\hline
$a$ ($\times10^{-4}%
\operatorname{K}%
^{-1}$) & $2.1$ \cite{Chen} & $2$ \cite{Chen} & $0.7$ \cite{Tomeno}\\\hline
$g$ ($\times10^{-16}%
\operatorname{cm}%
^{2}$) & $1$ \cite{Blinz and Zeks} & $1$ \cite{Blinz and Zeks} & $0.3$
\cite{Scrymgeour}\\\hline
$K_{IC}$ ($\times10^{8}%
\operatorname{erg}%
\operatorname{cm}%
^{-5/2}$) & $0.6\div2.0$ \cite{BaTiO3Toughness} & $1.4$ \cite{Jones and
Hoffman} & $1$ \cite{Shi}\\\hline
$\kappa$ ($\times10^{-14}%
\operatorname{s}%
$) & $1$ \cite{KappaBa} & $1$ \cite{kappaPb} & -\\\hline
$T_{m}$($%
\operatorname{K}%
$)$\ $ & $1898$ \cite{BaTiO3melting} & $1443$ \cite{PbTiO3melting} & $1526$
\cite{LiNbO3melting}\\\hline
$T_{0}$($%
\operatorname{K}%
$)$\ $ &
\begin{tabular}
[c]{ccc}%
II-III & III-IV & IV-V\\
$394$ & $284$ & $200$%
\end{tabular}
\cite{Tonkov} & $850$ \cite{Tonkov} & $1460$ \cite{Lehnert}\\\hline
$k$ ($\times10^{-8}%
\operatorname{K}%
\operatorname{cm}%
^{3}%
\operatorname{erg}%
^{-1}$) &
\begin{tabular}
[c]{ccc}%
$-0.8$ & $-0.3$ & $-0.1$%
\end{tabular}
\cite{Tonkov} & $-1$ \cite{Jabarov} & $0.018$ \cite{Scrymgeour}\\\hline
\end{tabular}
\end{table}

Here $T_{m}$ is the melting point. In the case of BaTiO$_{3}$ and PbTiO$_{3} $
$T_{0}$ is the temperature of the first order transition, while for
LiNbO$_{3}$ this value yields the Curie point. Note tahe the scatter of the
$K_{IC} $ values for BaTiO$_{3}$ (Table 1) originates from experimental
results obtained on ceramics with different properties (such as porosity,
grain size, etc.). It gives rise to the corresponding spread of the estimates
for BaTiO$_{3}$ parameters summarized in Table 2. In the case of BaTiO$_{3}$
the last row yields the values of the phase transition line slopes
corresponding to each its transition: II-III, III-IV and IV-V,

The estimates following form the above material constants are collected in
Table 2.


\begin{table}
\caption{Estimates of the LPT characteristics}%
\begin{tabular}
[c]{|c|c|c|c|c|}\hline
& $\Delta T_{\ast}$ ($%
\operatorname{K}%
$) & $\Delta T_{\ast}/T_{m}$ & $L_{f}$ ($%
\operatorname{nm}%
$) & $V_{c}$ ($\times10^{6}%
\operatorname{cm}%
/%
\operatorname{s}%
$)\\\hline%
\begin{tabular}
[c]{c}%
BaTiO$_{3}$ II-III\\
BaTiO$_{3}$ III-IV\\
BaTiO$_{3}$ IV-V
\end{tabular}
&
\begin{tabular}
[c]{c}%
$10^{3}$ to $10^{4}$\\
$10^{2}$ to $10^{3}$\\
$10^{2}$ to $10^{3}$%
\end{tabular}
&
\begin{tabular}
[c]{c}%
$1$ to $10$\\
$0.1$ to $1$\\
$0.1$ to $1$%
\end{tabular}
&
\begin{tabular}
[c]{c}%
$0.1$ to $1$\\
$1$\\
$1$%
\end{tabular}
&
\begin{tabular}
[c]{c}%
$1$\\
$0.1$ to $1$\\
$0.1$ to $1$%
\end{tabular}
\\\hline
PbTiO$_{3}$ & $10^{3}$ & $1$ & $1$ & $1$\\\hline
LiNbO$_{3}$ & $30$ & $0.01$ & $10$ & -\\\hline
\end{tabular}
\end{table}


In BaTiO$_{3}$ the slopes of all phase transition lines are negative. For this
reason the transformation zones only show up above the corresponding
transition lines. Thus, during fracture of the cubic BaTiO$_{3}$ one will find
at the tip a zone containing the tetragonal phase III embedded into the cubic
matrix II. The above estimates show that this zone exists up to the
temperatures of at least $\sim100$ to $\sim1000%
\operatorname{K}%
$ above the temperature $T_{II-III}=394%
\operatorname{K}%
$ of the bulk phase transition.

The zone containing the orthorhombic phase IV embedded into the bulk
tetragonal phase III should be found at the crack tip taking place above the
bulk transition temperature $T_{III-IV}$. It should be especially noted that
the temperature interval of existence of the orthorhombic zone, $\Delta
T_{\ast III-IV}\sim100$ to $1000%
\operatorname{K}%
$, is greater than the "distance" between the transitions II-III and III-IV:
$T_{II-III}-T_{III-IV}=110%
\operatorname{K}
$. Thus, the orthorhombic zone shows up in the whole region of existence of
the tetragonal phase and will is detectable within the bulk cubic phase
together with the tetragonal zone. In other words, a two-phase transformation
zone should take place over the temperature $T_{II-III}$.

It is worth noting that the room temperature belongs to the temperature
interval in which the tetragonal bulk phase III exists. It makes observation
of the local phase transition phenomenon in the phase III of BaTiO$_{3}$ convenient.

Finally, the zone with the ferroelectric, rhomboedral phase V should take
place on the background of the ferroelectric, orthorhombic, bulk phase IV up
to $\sim100$ to $1000%
\operatorname{K}%
$ above the bulk transition temperature $T_{IV-V}$. One finds $T_{IV}%
-T_{V}=84$ $%
\operatorname{K}%
$ and $\Delta T_{\ast IV-V}\gtrsim T_{IV}-T_{V}$. This implies that the
rhombohedral zone should be detectable within the whole domain of existence of
the bulk, orthorhombic phase IV and as well as in the tetragonal matrix III. A
proper description of such multiphase zones is, however, complex; we leave it
for another paper.

Since in PbTiO$_{3}$ one finds $k<0$, it only exhibits the transformation zone
during fracture of the cubic phase (i.e. above the transition line of the
phase diagram). From its phase diagram \cite{Jabarov} one can see that at the
atmospheric pressure the cubic phase only exists at the temperature over about
$T_{0}\approx850%
\operatorname{K}%
$. If the PbTiO$_{3}$ fracture takes at temperatures between $T_{0}$ and about
$1000%
\operatorname{K}%
$ above $T_{0}$, our estimates predict that it will be followed by formation
of the tetragonal transformation zone at the crack tip embedded into the
matrix of the cubic phase.

It should be noted that both in the case of BaTiO$_{3}~$and PbTiO$_{3}$ the
ratio of the temperature shift, $\Delta T_{\ast}$, to the melting temperature,
$T_{m}$, is between $\sim0.1$ and $\sim1$. Since the whole phase diagram spans
between $0%
\operatorname{K}%
$ and $T_{m}$, this shows that the region of the zone existence covers a
considerable part of or even the whole phase diagram.

In principle, the slope, $k$, of the phase diagram line may have any value,
including a very large or a small one. The latter is the case of LiNbO$_{3}$
where it gives rise to a relatively small shift of $\sim10%
\operatorname{K}%
$.

\subsubsection{Typical values of the temperature shift $\Delta T_{\ast}$}

It should be noted that the huge values of $\Delta T_{\ast}\sim100$ to $1000%
\operatorname{K}%
$ are not only inherent for BaTiO$_{3}$, PbTiO$_{3}$. To argue that let us
note the expression (\ref{DeltaTMain}) for the temperature shift at $V=0$ can
be written as:
\begin{equation}
\Delta T_{\ast}\sim r_{c0}^{-2/3}(kK_{IC})^{4/3}\label{DeltaTshort}%
\end{equation}
where $r_{c0}=(g/a)^{1/2}$ is the order parameter correlation radius at the
"distance" of $1%
\operatorname{K}%
$ from the transition line. The latter represents one of the most accessible
parameters, since it can be extracted from the width of the X-ray spectrum
peaks \cite{Krivoglaz} typically exhibiting the value of $r_{c0}\sim1$ to $10$
$%
\operatorname{nm}%
\operatorname{K}%
^{1/2}$, as well as that there is a typical value of the slope of the phase
diagram line $k\sim1$ to $10$ $%
\operatorname{K}%
/%
\operatorname{kbar}%
\sim(0.1\div1)\times10^{-8}%
\operatorname{K}%
\operatorname{cm}%
^{3}%
\operatorname{erg}%
^{-1}$ \cite{Tonkov}, and the typical value of the fracture toughness of the
inorganic solids is $K_{IC}\sim1%
\operatorname{MPa}%
\operatorname{m}%
^{3/2}\sim10^{8}%
\operatorname{erg}%
\operatorname{cm}%
^{-5/2}$ \cite{CherepanovTables} one finds the typical temperature shift:%

\[
\Delta T_{\ast}\sim10^{2}\text{ to }10^{4}%
\operatorname{K}%
\]
Since the typical values of the melting point of inorganic solids is
$T_{m}\sim10^{3}%
\operatorname{K}%
$, one concludes that $\Delta T_{\ast}$ typically covers a considerable part,
if not the whole phase diagram above or below the line of the bulk phase transition.

\subsection{On difficulties and possibilities to detect a transformation
process zone outside of the hysteresis region}

In the introduction we listed a number of materials for which LPT observation
has been reported. A relatively small number of such materials others than
those of the martensite type seems to contradict our main findings. This is,
however, only an apparent contradiction. Detailed inspection of the works
cited above shows that in most of them the local phase transition has been
detected by the analysis of the fracture surface available after the sample
has been broken, the so-called, "post mortem" examination. In the case of the
zirconia, for example, the fracture surface exhibited a layer of the
monoclinic daughter phase about $1%
\operatorname{\mu m}%
$ thick on top of the tetragonal mother phase surviving a considerable time
after fracture. This requires the zirconia to be deep within the hysteresis
region of its phase diagram. Indeed, martensitic transformations in both
martensite-austenite metals and zirconia exhibit wide hysteresis regions. This
is, however, rare, for most solids the hysteresis does not exceed $\sim10%
\operatorname{K}%
$, while the second order transitions have no hysteresis at all.

In contrast to the local phase transition within the hysteresis, the
transformation zone outside of the hysteresis region is only present under
stress. As soon as the stress is removed it immediately disappears. It cannot,
therefore, be detected by the "post mortem" inspection of the fracture surface.

Further, at a high temperature one may observe no zone at $K_{I}=K_{IC}$, but
since $\Delta T_{\ast}\sim K_{I}^{4/3}$ the zone may show up at higher stress
intensity factor, that is, at the tip of a propagating, rather than motionless crack.

Detection of a local phase transition outside the hysteresis region is,
therefore, a challenging experimental task.

\subsection{The crack tip zone concept}

The notion "process zone" is already in use since long time. Its emergence
reflects the understanding that the small domain in the immediate vicinity of
the tip of a brittle crack should have special properties due to high stresses
present there. Behavior of propagating cracks exhibits pronounced deviations
from predictions of linear fracture mechanics \cite{Freund}. It is generally
believed that they can only be explained by mechanisms located within the
process zone \cite{Fineberg}, \cite{Field}. Nevertheless, the content of this
notion stays so far on an intuitive level. Below we outline the process zone
concept from the physical point of view.

Equation of motion of the crack tip can be regarded as that of balance of
forces, the driving force, $K_{I}^{2}/E$, being balanced by (i) $K_{IC}^{2}/E
$ representing a kind of "dry" friction force and (ii) by additional
resistance force generated within the process zone. Within the force balance
concept the zone is the source of resistance of the solid to the crack
propagation, since it is here that this additional force is generated.

The concept of the process zone itself implies that it is possible to
distinguish the solid inside from that outside of the zone using at least one
physical property. Though in principle one can imagine a zone with a smooth,
gradual variation of all its properties, we believe that the situation in
which at least one physical property of the solid abruptly varies across the
zone boundary is much more realistic and, hence, often met.

We parametrized the differences of the solid properties inside the zone from
those outside by the field, $\eta$. Without the loss of generality we assume
$\eta=0$ inside the zone, while vanishing outside. In this respect our point
of view is akin to the popular phase field approach \cite{PhaseField}. In
principle, this field may describe either an abrupt quantitative variation of
some solid property, or its qualitative change. In the latter case the
situation is usually qualified as a phase transition, while $\eta$ is referred
to as the order parameter. In the present paper we focus on this latter case.

At present the terms "phase transition" and "order parameter" unify a crystal
structure variation of solids with bifurcations in non-linear systems, both
equilibrium and non-equilibrium, such as e.g., bifurcations taking place
during chemical reactions \cite{Cross}. It is tightly related to the fact that
each of these transitions can be described by its inherent order parameter(s).
They exhibit a few classes of universalities generating the corresponding
types of kinetic equations imposed on the order parameter(s), therefore,
giving rise to different configurational forces acting on the zone boundary
and, hence, exerted on the crack tip. The most often met universality class is
associated with the order parameter obeying the Ginzburg-Landau-Khalatnikov
equation \cite{Cross}, since it is related to the simplest bifurcation. This
one is addressed in the present paper. The zone at the crack tip may, however,
be associated with any of such phase transitions or their combinations.

The long-wave elastic field at the tip of a brittle crack exhibits its own
universal behavior expressed by the well-known small-scale approximation for
the stress or strain field $\varepsilon_{ii}\sim r^{-1/2}$. This latter
universality combines with one of those mentioned above.

We illustrate the above ideas within the example of classical structural phase
transitions related to changes in lattice structures in crystals in response
to the temperature and/or pressure variations. Structural phase transitions
can be classified according to their own classes of universalities. These are
related first of all to the symmetry change taking place during the
transition(s) \cite{LandauStat}, \cite{Gufan}, \cite{Toledano}. The latter is
manifested in (i) the number of components of the order parameter and (ii)
structure of the Ginzburg-Landau-Khalatnikov equation. It, in particular,
defines the form of interaction of the order parameter with other degrees of
freedom including elastic ones.

One observes in addition that the ranges of the values of material parameters
in use are quite narrow. This allows one to determine typical numerical values
of the parameters derived in our model.

To conclude, within our approach the process zone is regarded as a domain
which is unambiguously different from the bulk of the solid. The difference
may either be quantitative (such as a prominent variation of at least one
physical parameter), or qualitative (e.g. the difference in its symmetry,
crystal structure or chemical composition). Such a variation (or their
combination) takes place due to the high stress in the vicinity of the crack
tip, as a consequence of the nonlinearity. Our approach is, further, based on
the accounting for the universalities inherent both to the fracture mechanics
and bifurcations of non-linear systems offering a way to classify zones
according to the universality classes as well as on the observation of their
typical features. The zones related to the elastic nonlinearity
\cite{Bouchbinder}, formation of the secondary cracks \cite{Fineberg}, crack
tip chemical reactions \cite{Rostom}, electronic structure variation
\cite{Boulbitch and Fisenko} as well as the transformation zone reported in
the present paper fit into this general scheme.

\section{Appendix A: Exclusion of the acoustic variables}

Below we exclude acoustic degrees of freedom from the equation of motion as it
has been proposed in the paper \cite{Korzhenevskii}. Making use of
(\ref{GLK1}, \ref{stress2}) one can express the displacement vector, $u_{i}$
as%
\begin{equation}
u_{i}=u_{i}^{(0)}(\mathbf{r})+A\int G_{ij}(\mathbf{r}-\mathbf{r}^{\prime
})\frac{\partial\eta^{2}(\mathbf{r}^{\prime})}{\partial x_{j}^{\prime}}%
d^{2}x^{\prime}\label{Green1}%
\end{equation}
where $u_{i}^{(0)}(\mathbf{r})$ is the displacement field created by the
"undressed" crack (that is, the crack without any LPT) and $G_{ij}%
(\mathbf{r})$ is the Green function of the elastic solid with a cut. To the
best of our knowledge the explicit form of such a Green function is unknown.
We approximate it with the Green function of the infinite
elastically-isotropic body \cite{LandauMech}. Passing to the reciprocal space
under the integral in the representation (\ref{Green1}) and making use of the
identity
\[
G_{ik}(\mathbf{q})q_{j}q_{k}=\frac{1-2\sigma}{2\mu(1-\sigma)}\frac{q_{i}q_{j}%
}{\mathbf{q}^{2}}%
\]
\ (where $G_{ik}(\mathbf{q})$ is the Fourier-transform of the Green function)
one finds the strain field, $\varepsilon_{ik}(\mathbf{r})$ in the following
form:
\begin{equation}
\varepsilon_{ik}(\mathbf{r})=\varepsilon_{ik}^{(0)}(\mathbf{r})+\frac
{A(1-2\sigma)}{2\mu(1-\sigma)}\int\frac{q_{i}q_{k}}{q^{2}}Q(\mathbf{q}%
)\exp(i\mathbf{qr})\frac{d^{3}q}{(2\pi)^{3}}\label{Green2}%
\end{equation}
where $\varepsilon_{ik}^{(0)}(\mathbf{r})$ is the strain of the "undressed"
crack and
\[
Q(\mathbf{q})=2\pi\delta(q_{z})\int\eta^{2}(x,y)\exp[i(q_{1}x+q_{2}y)dxdy
\]
is the Fourier-image of $\eta^{2}(\mathbf{r})$. The result (\ref{Green2})
leads one to the following expression for the trace of the strain field:%
\begin{equation}
\varepsilon_{ii}(\mathbf{r})=\varepsilon_{ii}^{(0)}(\mathbf{r})+\frac
{A(1-2\sigma)}{2\mu(1-\sigma)}\eta^{2}(\mathbf{r})\label{trace}%
\end{equation}
The first term in the right-hand sides of any of the expressions
(\ref{Green1}), (\ref{Green2}) and (\ref{trace}) describes the strain field
generated by the "undressed" crack, while the second one yields the LPT
contribution. Substitution of the strain trace (\ref{Green2}, \ref{trace})
into the free energy (\ref{FreeEnergy}, \ref{FreeEnergyDensity}, \ref{fch},
\ref{fel}) yields the effective free energy (\ref{EffectiveFreeEnergy}) with
the factor $\beta$ (\ref{beta}) in front of $\eta^{4}$ instead of $\beta_{0}$.

\section{Appendix B: Branching equation in the low-temperature phase}

In the low-temperature phase the branching equation is mostly convenient to
obtain starting from the effective free energy. The latter has the form:%

\begin{equation}
F_{\text{eff}}=F_{0}+I_{2}\left(  \alpha_{\ast2}-\alpha\right)  \xi_{2}%
^{2}+\frac{1}{3}I_{3}\xi_{2}^{3}+\frac{I_{4}\beta}{4}\xi_{2}^{4}%
\label{FeffLow}%
\end{equation}
It differs from the expression (\ref{Feff}) by the existence of the cubic
term, $I_{3}\xi^{3}$, where
\begin{equation}
I_{3}=\frac{1}{\sqrt{g\beta}}%
{\displaystyle\int}
\Psi_{\ast}^{3}(r)\left(  -\alpha\beta-\frac{\beta B\cos(\theta/2)}{\sqrt{r}%
}\right)  ^{1/2}rdrd\theta\label{I3}%
\end{equation}
Passing to dimensionless variables: $r=\rho/R_{2}$, $V=\nu V_{\text{c}}$ one
finds:
\begin{equation}
I_{3}(\nu_{2})=%
{\displaystyle\int}
A(\rho,\theta)B(\rho,\theta)\rho d\rho d\theta\label{I3zero}%
\end{equation}

where
\[
A(\rho,\theta)=\exp\left\{  3\times2^{2/3}\sqrt{\rho}\cos(\theta
/2)-3\times2^{2/3}\rho\left[  1+\nu\cos(\theta)\right]  \right\}
\]%
\[
B(\rho,\theta)=\frac{3}{2\times2^{1/6}}\left[  \left(  1-\nu_{2}^{2}\right)
\sqrt{\rho}-2\times2^{1/3}\cos(\theta/2)\right]  ^{1/2}%
\]
The integrals $I_{2}$ and $I_{4}$ are defined as in (\ref{In}).

The integrand of $I_{3}$ becomes complex as soon as the expression $\left(
1-\nu_{2}^{2}\right)  \sqrt{\rho}-2\times2^{1/3}\cos(\theta/2)$ under the
radical becomes negative. This takes place along the line
\begin{equation}
\rho_{0}(\theta)=\frac{4\times2^{2/3}\cos^{2}(\theta/2)}{(1-\nu_{2}^{2})^{2}%
}\label{Contour1}%
\end{equation}
where $\eta_{0}(\mathbf{r})$ (\ref{LowTemp}) turns into zero. At smaller
values of $r$ one finds $\eta=0$. For this reason in $I_{1,2,3}$ one should
only integrate over $\rho$ from $\rho_{0}(\theta)$ to infinity, while the
integration over $\theta$ runs from $-\pi$ to $\pi$.

The integration has been done numerically using a standard NIntegrate routine
of Mathematica 10.1 \cite{Wolfram} employing an even-odd subdivision method
with the local adaptive strategy. Below only the ratios $I_{2}I_{4}/I_{3}^{2}$
and $I_{3}/I_{4}$ are used. These ratios and their fitting by simple
functions:%
\begin{equation}
s_{1}(\nu)=I_{2}I_{4}/I_{3}^{2}\approx0.59+3.13\nu_{2}-9.50\nu_{2}^{2}%
+9.00\nu_{2}^{3}-3.23\nu_{2}^{4}\label{S1}%
\end{equation}%
\begin{equation}
s_{2}(\nu)=I_{3}/I_{4}\approx1.45-1.85\nu_{2}+2.44\nu_{2}^{2}+\frac
{0.31}{1-0.98\nu_{2}^{2}}\label{s2}%
\end{equation}

The effective free energy (\ref{FeffLow}) has a cubic term. Since it is
positive, one finds that the left minimum of the free energy (\ref{FeffLow})
is more pronounced. Analogously to the static state one concludes that the
solution of the branching equation should be chosen that corresponds to this
deeper minimum, that is, the negative one. Let us note that in a general case
the sign of $\delta\eta$ is opposite to the one of $\eta_{0}$. Should we have
chosen a negative $\eta_{0}$, we will get the positive sign for $\delta\eta$.

This solution of the branching equation takes the form:%

\begin{equation}
\xi_{2}=-\frac{s_{2}(\nu_{2})B^{2/3}}{g^{1/6}\beta^{1/2}}\left\{
1+\sqrt{1+s_{1}(\nu_{2})\left[  1+2\frac{g^{1/3}}{B^{4/3}}\left(  \alpha
-\frac{\kappa^{2}V^{2}}{8g}\right)  \right]  }\right\} \label{Solution2}%
\end{equation}

\section{Appendix C. Simulation: technical details}

To perform simulations we used the software COMSOL 4b. Equations have been
simulated in a half-plane $y\geq0$. A semi-circular domain has been defined
with the diameter, $D=50$. By trial and error we find that it is large enough,
to let the solution vanish well far from the domain boundary. The initial mesh
size of $5$ has been chosen, but the adaptive mesh refinement option has been
further used to automatically refine the mesh as appropriate. The no-flux
boundary condition has been set at the boundary $y=0 $ and the condition $u=0$
at the rest of its boundary. A straightforward simulation of the static
equation (\ref{Sim1}) with such boundary conditions, however, only returns the
trivial solution $u=0$ at any value of the control parameter $q$. To avoid
this instead of (\ref{Sim1}) we introduced a pseudo-dynamic equation:
\begin{equation}
\frac{\partial u}{\partial t_{\text{ps}}}=\Delta_{1}u+2^{1/3}\nu\frac{\partial
u}{\partial x_{1}}-\left[  q-\frac{\left[  \left(  x_{1}^{2}+y_{1}^{2}\right)
^{1/2}+x_{1}\right]  ^{1/2}}{\left(  x_{1}^{2}+y_{1}^{2}+\varepsilon\right)
^{1/2}}\right]  u-u^{3}\label{PseudoTimeStepping}%
\end{equation}
where $u=u(x_{1},y_{1},t_{ps})$, $t_{\text{ps}}$ is the pseudo-time and
$\left[  \left(  x_{1}^{2}+y_{1}^{2}\right)  ^{1/2}+x_{1}\right]
^{1/2}/\left(  x_{1}^{2}+y_{1}^{2}+\varepsilon\right)  ^{1/2}$ is equal to
$\cos(\theta/2)/r_{1}^{1/2}$, regularized in the vicinity of the point
$r_{1}=0$ by a small parameter $\varepsilon=0.0001$. Stable solutions of the
static equation (\ref{Sim1}) represent fixed points of the dynamic system
(\ref{PseudoTimeStepping}). As initial condition we used a smoothed step
function, only unequal to zero in a vicinity of the point $(0,0)$.%

\begin{figure}
[ptb]
\begin{center}
\includegraphics[
height=2.1802in,
width=3.2655in
]%
{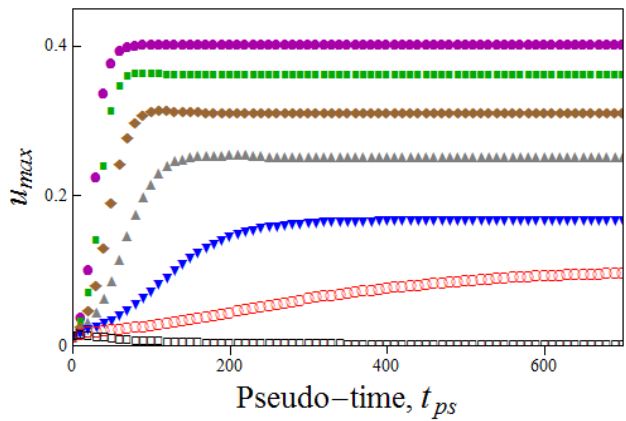}%
\caption{Illustration of the converging of the calculations with the
pseudo-time showing the convergence of the amplitudes of the rescaled order
parameter, $u_{\max}$, at different $q$ values. Filled disks: $q=0.3$, filled
squares: $0.32$, filled diamonds: $0.34 $, vertex-up triangles: $0.36$,
vertex-down triangles: $0.38$, open circles: $0.39$, empty squares: $0.41$.
Note that in the case of $q=0.39$ (point-down triangles) the $u_{\max}(t)$
dependence still exhibits a slope and a for the satisfactory convergence a
longer process was used (not shown).}%
\label{convergence}%
\end{center}
\end{figure}

The dynamic system has been solved using the direct MUMPS solver with the BDF
time stepping. The convergence of the solution to its fixed point has been
controlled by the behavior of the $u_{\text{max}}$, the maximum value of the
function $u(x_{1},y_{1},t_{ps})$. By trials we found that 700 pseudo-time
steps ensure a good convergence, though sometimes it has been necessary to
keep the process as long as 3000 steps. Figure \ref{convergence} shows the
example of such a convergence study for a number of simulations in which all
parameters except $q$ were fixed, while $q$ varied.

One can see that far from the bifurcation point the convergence takes place
well before $700$ pseudo-time steps are done. As it can be expected, the
situation is different in the close vicinity of the bifurcation ($q=0.38$ and
$0.39$ corresponding to the vertex-down triangles and open circles in Fig.
\ref{convergence}). Even here $700$ pseudo-time steps guarantee a rather
reliable convergence.


\begin{thebibliography}{99}
\bibitem {noteDissFunction}It only accounts for the dissipation due to the
evolution of the order parameter, rather than accoustic degrees of freedom. To
account for the dissipation of the latter one needs to add the corresponding
terms to the dissipation function. This is, however, not done in the present paper.
\end{thebibliography}
\end{document}